\documentclass[pre,twocolumn,showpacs]{revtex4}
\usepackage{graphicx}
\begin{document}
\title{Tagged Particle Correlations in the Asymmetric Simple Exclusion 
Process : Finite Size Effects}
\author{Shamik Gupta$^1$, Satya N. Majumdar$^2$, Claude
Godr\`eche$^3$, and Mustansir Barma$^1$
}

\affiliation{%
$^1$Department of Theoretical Physics, Tata Institute of Fundamental Research, Homi Bhabha Road, Mumbai 400 005, India\\
$^2$ Laboratoire de Physique Th\'eorique et Mod\`eles Statistiques,
Universit\'e
Paris-Sud, Orsay-Cedex, France\\
$^3$Service de Physique Th\'eorique, CEA Saclay, France \\
}%
\date{\today}
\begin{abstract}
We study finite size effects in the variance of the displacement of a
tagged particle in the stationary state of the Asymmetric Simple Exclusion Process (ASEP) on a ring of  size $L$. The
process involves hard core particles undergoing stochastic driven
dynamics on a lattice. The variance of the displacement of the tagged particle, averaged with
respect to an initial stationary ensemble and stochastic evolution,  grows
linearly with time at both small and very large times.  We find that at
intermediate times, it shows oscillations with a well defined size-dependent
period. These oscillations arise from sliding density fluctuations (SDF) in the stationary state with respect to the drift of the tagged particle, the density
fluctuations being transported through the system by kinematic waves. In the
general context of driven diffusive systems, both the Edwards-Wilkinson (EW)
and the Kardar-Parisi-Zhang (KPZ) fixed points are unstable with respect to the
SDF fixed point, a flow towards which is generated on adding a gradient term to the EW and the KPZ time-evolution equation. We also study tagged particle
correlations for a fixed initial configuration, drawn from the stationary
ensemble, following earlier work by van Beijeren. We find that the time dependence of this correlation is determined by the
dissipation of the density fluctuations. We show that an exactly
solvable linearized model captures the essential qualitative features
seen in the finite size effects of the tagged particle correlations in the
ASEP.  Moreover, this linearized model also provides an exact
coarse-grained description of two other microscopic models.
\end{abstract}
\pacs{02.50.Ey, 05.40.-a, 05.60.Cd, 66.30.Dn, 05.50.+q}
\maketitle
\section{Introduction}
The Asymmetric Simple Exclusion Process (ASEP) is a prototype model to
study driven diffusive motion \cite{ligget}. The process involves
hard-core particles hopping between neighboring sites of a lattice, with
an asymmetry in hopping rates which incorporates the effect of an
external drive. In the case of completely asymmetric dynamics, the ASEP
reduces to the totally asymmetric simple exclusion process (TASEP). In
either case, the system, on a periodic
lattice, settles at long times into a
non-equilibrium stationary state in which each particle has an average
drift velocity, and there is a finite current through the system.
\begin{figure}[h!]
\begin{center}
\includegraphics[scale=0.7]{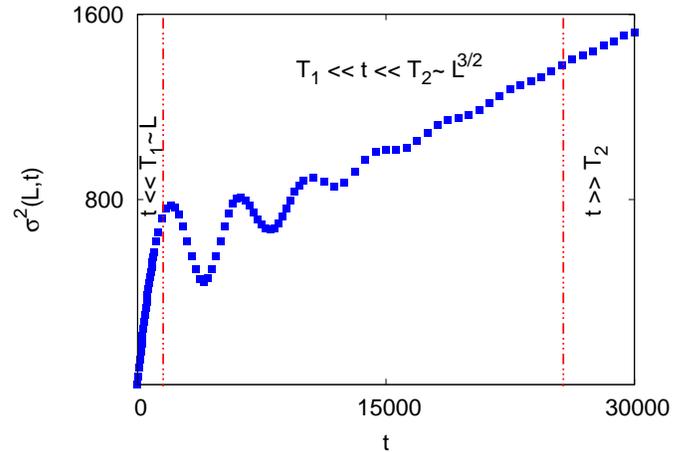}
\caption{(Color online) Monte Carlo (MC) simulation results for the time
dependence of the variance $ \sigma^{2}(L,t)$ for the
TASEP on a finite lattice, averaged over $10^{5}$ MC runs. The different time regimes are also marked.
Here,
$L=1024$, the particle density $\rho=0.25$.}
\label{1024-256-tc}
\end{center}
\end{figure}

In this paper, we are interested in the temporal growth of the fluctuations in the
displacement of a tagged ASEP particle in the steady state in a
\textit{finite} system of size $L$ in one dimension. We monitor
the variance $\sigma^{2}(L,t)$ of a \textit{tagged} particle around its
average in time $t$, starting from the stationary ensemble of configurations.
Some asymptotic results are already known \cite{vanb1,De Masi,derrida1,MB1,MB3}. These are summarized below.  

(a) On taking the limit $L \rightarrow \infty$ first, followed by the limit
$t \rightarrow \infty$, the fluctuations are diffusive, growing linearly in time:
$\sigma^{2}(L,t)\sim D_{0}t$. Here, $D_{0}$ is a known function of the
particle density and the external bias \cite{De Masi}, and is given
later in the paper.

(b) In the opposite limit with $t \rightarrow \infty$
before $L \rightarrow \infty$,
$\sigma^{2}(L,t)$ again behaves diffusively, i.e., $\sigma^{2}(L,t)\sim
D(L)t$, but now the diffusion constant
$D(L)$ depends on the system size $L$ and scales as $D(L)\sim
\frac{1}{\sqrt{L}}$ for large $L$ \cite{derrida1}. A natural question arises: What is the
full behavior of $\sigma^{2}(L,t)$ as a function of time in between
these
two extremes? In this paper, we show that in the intermediate regime, $\sigma^{2}(L,t)$
shows striking oscillations as a function of time. Indeed, there are two
time-scales in the problem, $T_{1} \sim L$ and $T_{2} \sim L^{3/2}$. The
limiting behaviors (a) and (b) hold in the regimes $t \ll T_{1}$ and $t \gg
T_{2}$, respectively, and in between ($T_{1} \ll t \ll T_{2}$), the quantity
$\sigma^{2}(L,t)$ oscillates with time, as shown in Fig. \ref{1024-256-tc}.

In measuring the fluctuations, if one does not start from the 
stationary ensemble, but instead from an
arbitrary but \textit{fixed} configuration drawn from the stationary
ensemble, the corresponding variance $s^{2}(L,t)$ of the displacement behaves very
differently from $\sigma^{2}(L,t)$. The behavior of $s^{2}(L,t)$, in the
limit of an infinite system, is known: $s^{2}(t)\equiv \lim_{L
\rightarrow \infty}s^{2}(L,t)\sim t^{2/3}$ \cite{vanb1}. For
a finite system, we find a similar growth until a characteristic time $T^{*}\sim
L^{3/2}$, after which this correlation grows linearly in time: $s^{2}(L,t) \sim D(L)t$ with $D(L) \sim \frac{1}{\sqrt{L}}$. 

Some of the results for the tagged particle correlations discussed in
this paper were known previously. However, a complete
and a consistent picture on a
finite lattice is lacking in the literature. The principal purpose of this
paper is to discuss finite size effects on the tagged
particle correlation in the ASEP on a one-dimensional ring from a unified point of view. In order
to do this, we include some known results for completeness, in addition
to the new results on the finite size effects. The main results are summarized below. 

{\bf{1.}} The behavior of $\sigma^{2}(L,t)$ as a function of time can be characterized by two time scales $T_{1} \sim L$
and $T_{2} \sim L^{3/2}$, as shown in Fig. \ref{1024-256-tc}.

(a) In the initial linear regime ($t \ll T_{1}$), the variance
$\sigma^{2}(L,t) \sim D_{0}t$ \cite{sepdis}.

(b) In the oscillatory regime ($T_{1} \ll t \ll T_{2}$), the quantity $\sigma^{2}(L,t)$ oscillates as a function 
of $t$. The amplitude of oscillations is proportional to the system size
$L$ while the time period of 
oscillations is $T=L/u$, where $u$ is the mean velocity of the stationary state density
fluctuations relative to the average drift velocity of the 
particles. 

(c) In the late time regime ($t \gg T_{2}$), the quantity $\sigma^{2}(L,t) \sim
D(L)t$ with $D(L) \sim \frac{1}{\sqrt{L}}$.

\begin{figure}[h!]
\begin{center}
\begin{tabular}{cc}
{\includegraphics[scale=0.7]{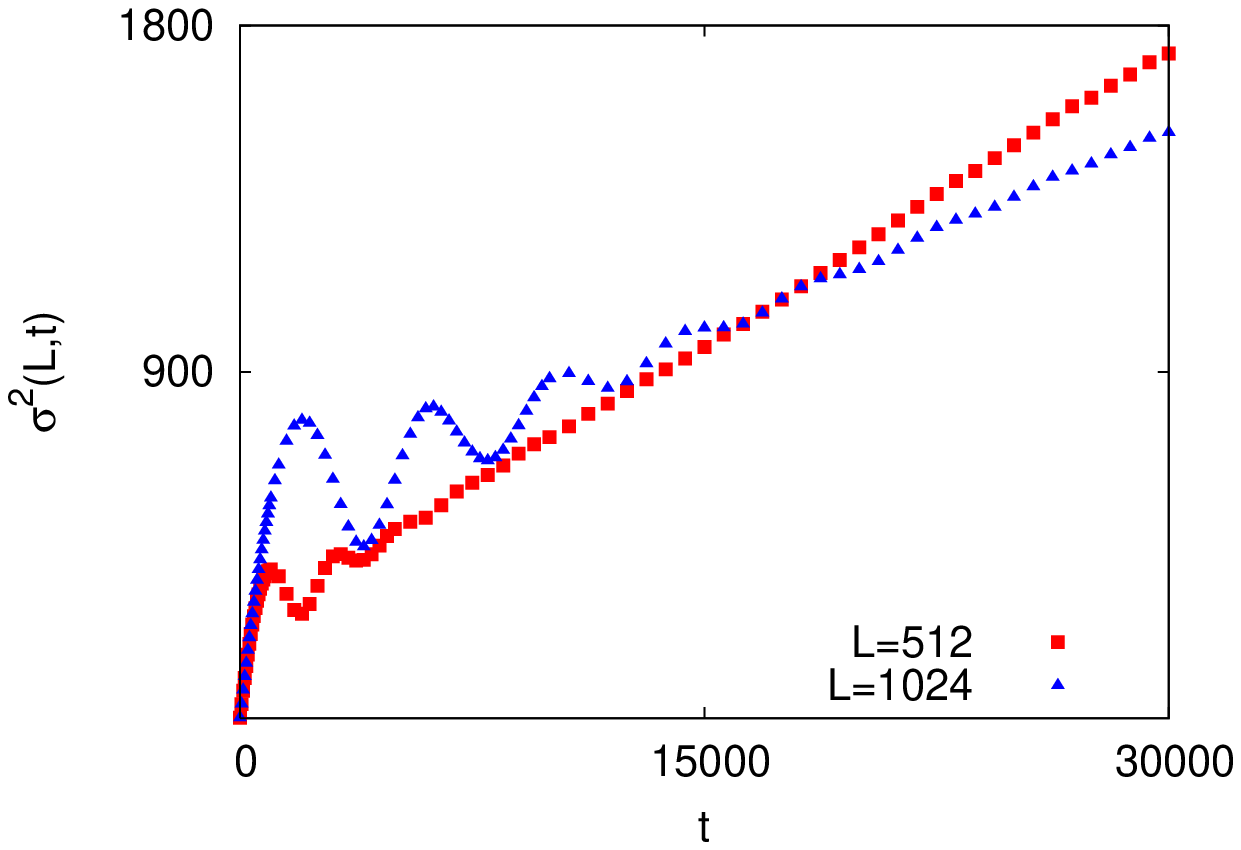}}\\
{\includegraphics[scale=0.7]{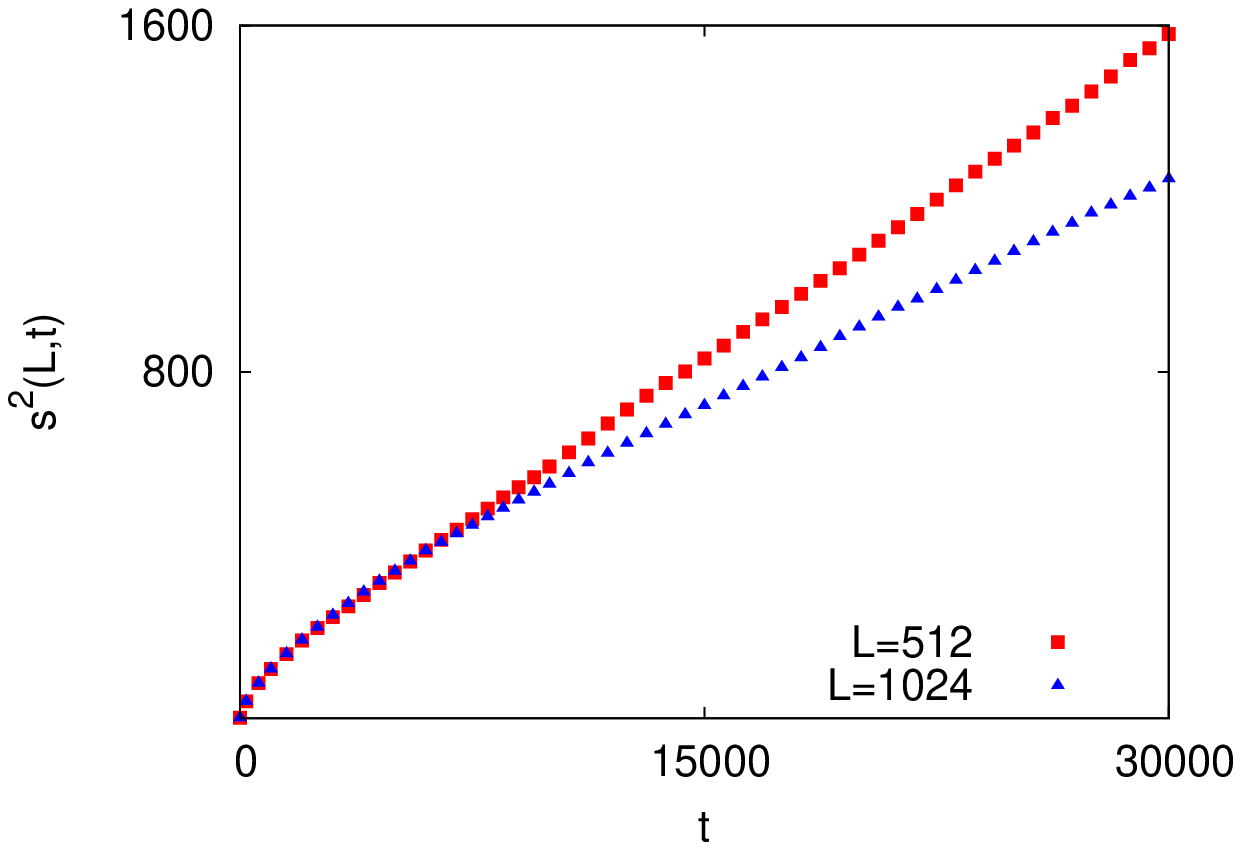}}
\end{tabular}
\caption{(Color online) MC simulation results for the time 
dependence of the variance of the displacement for the TASEP on a finite lattice for
$ \sigma^{2}(L,t)$ and $ s^{2}(L,t)$. In both cases, the
averaging is over $10^{5}$ MC runs. Here, the density $\rho=0.25$. The two system sizes are $512$ and $1024$.}
\label{tcvanb1024-512}
\end{center}
\end{figure}

{\bf{2.}} The time variation of $s^{2}(L,t)$ can be characterized by a
single time scale $T^{*} \sim L^{3/2}$. The
variation of $s^{2}(L,t)$ with time for two different system sizes is shown in Fig. \ref{tcvanb1024-512}.

(a) $t \ll T^{*}$: In this regime, $s^{2}(L,t) 
\sim t^{2/3}$.

(b) $t \gg T^{*}$. Here, $s^{2}(L,t) \sim D(L)t$ with $D(L) \sim
\frac{1}{\sqrt{L}}$.

The long-time behavior of both the functions $\sigma^{2}(L,t)$ and
$s^{2}(L,t)$ is diffusive, with the same diffusion constant $D(L)$ which
scales with the system size as $D(L) \sim \frac{1}{\sqrt{L}}$. This
behavior is attributed to the motion of the center-of-mass. 

The physical origin of oscillations in $\sigma^{2}(L,t)$ is the
occurrence of kinematic waves that transport stationary state density
fluctuations through the system (Appendix \ref{kinwaves}). Kinematic waves are known to arise in a variety of circumstances involving flow
e.g., in flood movement in long rivers \cite{lighthill1}, traffic
\cite{lighthill2,debch}, flow of granular particles through vertical
tubes and hoppers \cite{jysoolee}, motion of transverse fluctuations 
in interfaces \cite{MB2,MB5}, field-induced transport in random media 
as in drop-push dynamics \cite{ram1}. In our system, the velocity of the 
sliding density fluctuations (SDF) relative to the average drift of the
ASEP particle determines the period of the oscillations. Eventually, the
oscillations damp down as the density fluctuations dissipate owing to the stochasticity and
nonlinearity inherent in the dynamics. This dissipation in time is
well captured by the temporal behavior of the function $s^{2}(L,t)$, as we
show in this paper.

{\bf{3.}} The ASEP can be mapped to a non-equilibrium model of interface growth
in the Kardar-Parisi-Zhang (KPZ) universality class \cite{kpz}. The resultant time
evolution equation for the interface is the usual KPZ equation, augmented by a drift term which accounts for
the SDF. In this paper, we consider the interface equation with the
drift term in the linear approximation, and solve exactly for
$\sigma^{2}(L,t)$ and $s^{2}(L,t)$. Our analytical solution of the
linearized model captures the essential
qualitative features seen in these two quantities for the ASEP.
Moreover, we show that the linearized model is an exact coarse-grained
description of two specific microscopic models of interacting particles,
the Katz-Lebowitz-Spohn (KLS) model \cite{KLS} at a specific value of the temperature and
the Asymmetric Random Average Process (ARAP) \cite{arap}. Our analytic
solution quantitatively describes the behavior of the tagged particle
correlations for these two
models.

The paper is organized as follows. In Section \ref{themodel}, we define
the model and the variances $\sigma^{2}(L,t)$ and $s^{2}(L,t)$ of the
tagged particle process, followed by a discussion in the context of the KPZ interface equivalent to the ASEP. In
Section \ref{repinfsys}, we give a representation for the tagged
particle displacement on an infinite lattice, and show how it correctly accounts
for the known results for the variance. This is
followed by Section \ref{physicalarg} where we adduce physical arguments
for the observed behavior of $\sigma^{2}(L,t)$ and $s^{2}(L,t)$, both on
an infinite as well as a finite lattice. The time-evolution equation for the interface-equivalent
of the ASEP is solved
exactly in the linear approximation in Section \ref{exactsol} which
explains some qualitative features of the tagged particle correlations for the ASEP. As we show in
Section \ref{mappinglin}, the exact solution also explains
quantitatively the variance of the tagged particle displacement
for the KLS model at a specific value of the temperature and for the ARAP.
In Section \ref{com}, we discuss the center-of-mass
motion for the ASEP, followed by conclusions in Section \ref{summary}.

\section{Tagged particle correlations}
\label{themodel}
\subsection{The Model}
\label{themodel1}
We consider the ASEP on a periodic lattice of $L$ sites. $N$ 
indistinguishable hard core particles are distributed over the lattice
sites with each site either singly-occupied or empty. The particle
density $\rho=\frac{N}{L}$ is held
constant when the limit $L \rightarrow \infty, N \rightarrow \infty$ is
taken. Let $n_{i}=0,1$ with $i=1,2,\ldots,L$ denote the
occupancy of the $i$-th site. The
system evolves according to a stochastic dynamics: during an
infinitesimal time interval $dt$, a particle attempts to hop to the 
site to its right with probability $pdt$, to the left neighboring
site with probability $qdt$, and continues to occupy the original
site with probability $1-(p+q)dt$. The attempted hop is successful
only if the sought site is empty before the hop. $L$ attempted hops 
constitute one time step. Clearly the total number of particles is
conserved under the dynamics. For the TASEP, the motion of the particles is entirely in
one direction i.e., $p=1$ and $q=0$ or, vice-versa. Note that with $p=q$, each particle 
moves symmetrically to the left and to the right. The model then 
reduces to the symmetric Simple Exclusion Process (SEP), an 
equilibrium model of hard core particles diffusing on a lattice.   

It is evident from the rule of evolution that the
dynamics is ergodic. In the limit $t \rightarrow \infty$, the system settles
into a nonequilibrium stationary state. Ergodicity ensures uniqueness
of the stationary state. On a lattice of size $L$, the stationary
state is one in which all configurations $\mathbf{C}$ have the same
weight $p(\mathbf{C})=\frac{N!(L-N)!}{L!}$ \cite{ligget}. It then
follows that the two-point correlation function in the stationary 
state $\overline{n_{i}n_{j}} =\frac{N(N-1)}{L(L-1)}$ for
$i \ne j$. Here, as in the rest of the paper, the overbar is used
to denote an average over the stationary ensemble of 
configurations. In the thermodynamic limit, the steady state has a
product measure form. In the 
stationary state, the ASEP supports a steady current of particles 
whose mean value $J = (p-q)\overline{n_{i}(1-n_{i+1})}$ for a 
system of size $L$ is given by
$J=(p-q)\rho(1-\rho)+O(\frac{1}{L})$.
Correspondingly, the mean velocity of a particle in the
stationary state, given by $v_{P}=\frac{J}{\rho}$, equals
$v_{P}=(p-q)(1-\rho)+O(\frac{1}{L})$. 

\begin{figure}[h!]
\includegraphics[scale=0.7]{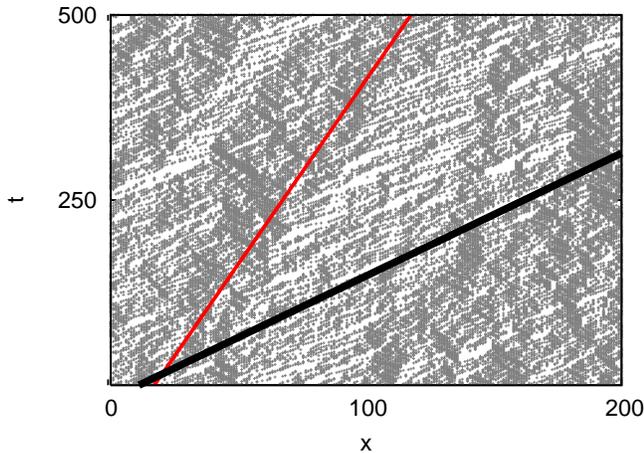}
\caption{(Color online) Time evolution for a stationary TASEP 
configuration. Here, the system size $L=200$, the density $\rho=0.4$.  
The inverse of the slope of the thick black line is the velocity of the tagged particle
$v_{P}\approx(p-q)(1-\rho)=0.6$ while that of the thin red line is the velocity of the density fluctuations
$v_{K}\approx(p-q)(1-2\rho)=0.2$. }
\label{timeev}
\end{figure}

Besides particle motion, there is
also a motion associated with the coarse-grained density fluctuations in the
stationary state of the ASEP. The density fluctuations are transported
as a kinematic wave with velocity $v_{K}=\frac{\partial J}{\partial \rho}$
(see Appendix \ref{kinwaves}). Thus,
relative to the average drift of the particles, the density
fluctuations `slide' with velocity $v_{K}-v_{P}$. We refer to this relative
motion as the sliding density fluctuations (SDF). Stochasticity and
nonlinearity in the dynamics lead to dissipation of the density
profile so that the wave of fluctuations ultimately dies down in time. The kinematic wave velocity for an infinite system is given by
$v_{K}=(p-q)(1-2\rho)$. This velocity, derived on the basis of
hydrodynamic considerations discussed in Appendix \ref{kinwaves}, also appears as the
imaginary part of the low-lying eigenvalues of the Markov matrix
governing the time evolution of the ASEP in the totally asymmetric case
$q=0$ \cite{kim}. Figure \ref{timeev} shows the time evolution of
a stationary ASEP configuration. One clearly sees that
the space-time trajectory of a particle is distinctly different
from that for a coarse-grained density fluctuation. 
\subsection{Two different observables}
The primary quantity of interest in this paper is the variance of the
displacement of a tagged particle around its mean trajectory. Note that
the displacement $y(n,t)$ of the $n$-th particle in time $t$ ranges from $-\infty$ to $\infty$ even though the system size $L$ is finite.

\subsubsection{Average over histories and initial conditions: 
$\sigma^{2}(L,t)$}
Here we start with the
stationary ensemble of the ASEP configurations at $t=0$ and monitor the motion of a tagged
particle. Let $\Delta_{n}(t)$ denote the deviation in the displacement
of the $n$-th
tagged particle from its average value in time $t$, the averaging being
with respect to both the initial stationary ensemble at $t=0$ and
stochasticity in the evolution of configurations. In symbols, 
\begin{eqnarray}
\Delta_{n}(t)\equiv y(n,t)-y(n,0)-\overline{\langle [y(n,t)-y(n,0)]
\rangle} \nonumber \\
~~~~\left[\mathrm{Ensemble ~ of ~ Initial ~ Conditions}\right].
\label{Deltadef}
\end{eqnarray}
The angular brackets denote averaging over the stochasticity in
the  evolution of configurations, while the overbar is used to denote
averaging with respect to the initial stationary ensemble at $t=0$. In the function $\Delta_{n}(t)$, the quantity 
$\overline {\langle y(n,t)-y(n,0)\rangle}$ gives the average 
displacement of the $n$-th tagged particle in time $t$ in the 
stationary state, and hence is equal to $v_{P}t$.  The variance of the 
displacement, measuring the stationary state fluctuations in
$\Delta_{n}(t)$, is given 
by 
\begin{equation}
\sigma^{2}(L,t) \equiv \overline{\langle \Delta_{n}^{2}(t) \rangle}.
\label{sigmadef}
\end{equation}
Since the system is translationally invariant, the above quantity is the
same for all particles, and hence we do an additional averaging of
$\sigma^{2}(L,t)$ over the particles.

 In this paper, we study how the finiteness of the system size $L$ 
 affects the behavior of $\sigma^{2}(L,t)$ in time. 
\subsubsection{Average over histories only: $s^{2}(L,t)$}
 Beginning with an arbitrary but \textit{fixed} configuration, chosen from the
 stationary ensemble and following its evolution in time, it is interesting to monitor the fluctuations in the displacement of a
 tagged particle, and study how finite size effects come into play here. These
 fluctuations were first studied by van Beijeren for an infinite
 system in \cite{vanb1}. In this case, one averages only over the stochasticity in the evolution of the configuration.
 We define $d_{n}(t)$ to be the deviation in the displacement of the
 $n$-th tagged particle from its mean value in time $t$, starting from an
 arbitrary but fixed configuration, drawn from the stationary ensemble
 at $t=0$. Explicitly, 
\begin{eqnarray}
d_{n}(t)\equiv y(n,t)-y(n,0)-\langle [y(n,t)-y(n,0)] \rangle \nonumber \\  
~~~~\left[\mathrm{Fixed ~ Initial ~ Condition}\right].
\label{ddef}
\end{eqnarray}
 The corresponding variance is given by
\begin{equation}
s^{2}(L,t)\equiv \langle d_{n}^{2}(t)\rangle.
\label{sdef}
\end{equation}
The system being translationally invariant, the above quantity is also averaged over
the particles.

We note that the behavior of both $\sigma^{2}(L,t)$ and
$s^{2}(L,t)$ can be extracted from two different limiting
values of a single function that measures the fluctuation in the 
displacement of a tagged particle in the stationary state. At time
$t=0$, we start with an arbitrary but fixed configuration
$\mathbf{C}_{0}$, drawn from the stationary ensemble of configurations.
After $\mathbf{C}_{0}$ has evolved in time for an interval $t_{0}$, we
start measuring the variance $C_{L}(t_{0},t_{0}+t)$ of the displacement of
a tagged particle around its average:  
\begin{widetext}
\begin{equation}
C_{L}(t_{0},t_{0}+t) \equiv \langle[y(n,t_{0}+t)-y(n,t_{0})-
\langle[y(n,t_{0}+t)-y(n,t_{0})]\rangle]^{2}\rangle,
\label{cldef}
\end{equation}
\end{widetext}
where the angular brackets denote averaging with respect to stochastic
evolution.

$(1)$ On taking the limit $t_{0} \rightarrow 0$, when the
averaging is only with respect to stochastic evolution, we get the 
function $s^{2}(L,t)$. 

$(2)$ On the other hand, since the system is ergodic, in the limit
$t_{0} \rightarrow \infty$, the
initial configuration $\mathbf{C}_{0}$ evolves into an ensemble of stationary configurations. Thus, the function $\lim_{t_{0} 
\rightarrow \infty}C_{L}(t_{0},t_{0}+t)$ measures fluctuations in the
stationary state where the averaging is with respect to both the
initial stationary ensemble and stochastic evolution. Hence, this 
quantity is identical to $\sigma^{2}(L,t)$.
\subsection{Interface equivalent}
Many of our results for $\sigma^{2}(L,t)$ and $s^{2}(L,t)$ can be
explained in terms of the continuum description of the ASEP, achieved by mapping it to a model of nonequilibrium
interface growth \cite{MB1} that belongs to the KPZ universality 
class of nonequilibrium interface dynamics \cite{kpz}.
The mapping that we use in this paper employs the tagging process of
the particles in a direct and essential way in the translation
\cite{note}. Let the particles be labeled $1,2,..,N$ sequentially at the initial
instant. Since the particle motion is in one dimension, and there is no overtaking, 
this labeling will be preserved for all
subsequent times. The corresponding interface is obtained by
identifying the tag label $n$ with the horizontal coordinate $n$ for
the interface, while the set $\{y(n,t)\}$, denoting the particle
locations at time $t$, maps onto the set of local interface heights $\{h(n,t)\}$ 
defined by
\begin{equation}
h(n,t)=y(n,t)-\frac{n}{\rho}.
\end{equation}
Since the inequality $y(n+1,t) \ge y(n,t)+1$ holds, it follows that the
interface heights satisfy $h(n+1,t) \ge h(n,t)+1-\frac{1}{\rho}$.
Periodic boundary condition implies $y(n\pm N)=y(n)\pm L$, and
correspondingly, $h(n \pm N,t)=h(n,t)$.

The dynamics of the interface involves the following moves in a small
time interval $dt$: the move $h(n,t) \rightarrow h(n,t+dt)=h(n,t)+ 1$ 
occurs with probability $pdt$ while the move $h(n,t)
\rightarrow h(n,t+dt)=h(n,t)- 1$ takes place with probability $qdt$. The
interface height remains the same with probability $1-(p+q)dt$. The attempt to increase, respectively
decrease, succeeds only if $y(n+1,t)-y(n,t) > 1$, respectively
$y(n,t)-y(n-1,t) > 1$. 
\begin{table}[h!]
\begin{center}
\begin{tabular}{|c|c|} \hline
ASEP & Lattice Interface Model\\ \hline
Particle Label $n$  & Spatial coordinate $n$ \\ \hline
Displacement $y(n,t)$ & Height $h(n,t)+\frac{n}{\rho}$ \\ \hline
Mean Velocity $v_{P}$ & Mean growth rate $\frac{\partial \langle h 
\rangle}{ \partial t}$ \\ \hline
\end{tabular}
\caption{Mapping ASEP to the interface.}
\label{mapping}
\end{center}
\end{table}

Following the above prescription, the ASEP with $N$ particles and $L$ sites
maps onto a lattice model of the interface of length $N$. In order to get the continuum equation for the interface, we (i)
coarse-grain the particle labels so that the discrete tag label $n$ becomes the
continuous tag variable $x$, and (ii) divide $x$ by the particle density
$\rho$ to make $x$ into a spatial variable running between $0$ and $L$. The equation of motion of the interface, to lowest order of nonlinearity, is given by
\begin{eqnarray}
\frac{\partial h(x,t)}{ \partial t}=v_{P}+\Gamma 
\frac{\partial^{2}h}{\partial x^{2}}+u\frac{\partial h}{\partial x}+
\frac{\lambda}{2}\left(\frac{\partial h}{\partial x}\right)^{2}+\eta(x,t),
\label{kpzdrift}
\end{eqnarray}
where $v_{P}=(p-q)(1-\rho), \Gamma=\frac{1}{2}, u=\rho(p-q), \lambda =
-2\rho(p-q)$. Here, $\eta(x,t)$ represents a Gaussian noise with 
$\langle \eta(x,t) \rangle=0, \langle \eta(x,t) \eta(x',t') 
\rangle =2A\delta (x-x')\delta (t-t')$, where
$A=\frac{1}{2}\left(\frac{1-\rho}{\rho}\right)$. The boundary condition $h(n \pm
N,t)=h(n,t)$ on the height variable for the discrete interface now reads
$h(x\pm L,t)=h(x,t)$ in the continuum. The derivation of Eq.
\ref{kpzdrift} is relegated to the Appendix \ref{eqderivation}. The 
constant term $v_{P}$ and the drift term $u\frac{\partial h}{\partial
x}$ in Eq. \ref{kpzdrift} can be eliminated by a boost and a Galilean shift, respectively, as explained
below. In that case, Eq. \ref{kpzdrift} describes the 
time evolution equation of a KPZ interface \cite{kpz}. Note that Eq. \ref{kpzdrift} is not an exact
description of the time evolution for the ASEP density profile. It is
rather a coarse-grained description of the ASEP. One expects that the scaling properties of the correlation functions for both the ASEP
and its interface-equivalent are governed by the same KPZ fixed point,
and hence, are described by the same critical exponents and scaling
functions.  

For the symmetric exclusion process (SEP), the corresponding 
interface is an equilibrium one that does not move bodily. Rather, it fluctuates
about a stationary profile, and its time evolution is governed by
the Edwards-Wilkinson (EW) equation \cite{ew}, obtained by setting
$p=q$, implying $v_{P}=0,u=0,\lambda=0$, so that
\begin{equation}
\frac{\partial h(x,t)}{ \partial t}=\Gamma \frac{\partial^{2}h}
{\partial x^{2}}+\eta(x,t).
\label{ew}
\end{equation}

On including all the higher order nonlinearities, Eq. \ref{kpzdrift}
is replaced by the general form
\begin{equation}
\frac{\partial h(x,t)}{\partial t}=\Gamma \frac{\partial^{2}h}
{\partial x^{2}}+\sum_{m=0}^{\infty}\lambda_{m}\left(\frac{\partial h}
{\partial x}\right)^{m}+\eta(x,t),
\label{genint}
\end{equation}
where, e.g., $\lambda_{0}=v_{P}, \lambda_{1}=u, \lambda_{2}=
\lambda/2$, etc. The different coefficients $\lambda_{m}$ in Eq. \ref{genint} 
determine the scaling properties of the height fluctuations, as discussed
below. $\lambda_{0}$ can be eliminated by redefining the height variable $h \rightarrow
h'=h+\lambda_{0}t$.  The first-order gradient term,
$\lambda_{1}\frac{\partial h}{\partial x}$, can be eliminated from
Eq. \ref{genint} by Galilean shift $x \rightarrow x'=x+\lambda_{1}t$.

The stationary state height fluctuations are measured by
\begin{eqnarray}
S(x,t)&=&\overline{\langle
\left[h(x+x',t)-h(x',0)\right]^{2}}\rangle\nonumber \\
&-&\left[\overline{\langle h(x+x',t)-h(x',0)}\rangle\right]^{2},
\end{eqnarray}
where the angular brackets denote averaging over the stochastic
evolution of the interface, while the overbar represents averaging over
the initial stationary ensemble of the interface configurations.

If $\lambda_{1}$ is nonzero, the autocorrelation
function grows as $S(0,t) \sim t$ for large $t$ \cite{MB1,MB2,MB3}.
This can be explained in terms of sliding density fluctuations (SDF)
with respect to the particle motion, as discussed in Section 
\ref{physicalarg}. If $\lambda_{0}=\lambda_{1}=0$, the function $S(x,t)$ assumes the
scaling form
\begin{equation}
S(x,t) \sim t^{2\beta}Y\left(\frac{x}{t^{1/z}}\right)
\end{equation}
in the asymptotic limit $x,t \rightarrow \infty$, with $x/t^{1/z} =$
constant. Here, $\beta$ is the growth exponent
while $z$ is the dynamical exponent. These
exponents as well as the scaling function $Y$ are the same for all 
systems belonging to the same universality class.
The scaling function $Y(s)$ has the property that $Y(s) \rightarrow$
constant as $s \rightarrow 0$, implying that the height autocorrelation
$S(0,t) \sim t^{2\beta}$.  Further, as $s \rightarrow
\infty$, the function $Y(s) \rightarrow s^{2\alpha}$. Here, $\alpha$ is
the critical exponent related to $z$ and $\beta$ through
$z=\alpha/\beta$ \cite{barabasi}. It determines the roughness of the interface, $S(x,0) \sim
x^{2\alpha}$.

From a perturbative renormalization group (RG) analysis of Eq.
\ref{genint}, it is known that there are two distinct fixed points
in the parameter space. These two fixed points define two universality
classes. The first class is the one in which all $\lambda_{m} = 0$.
In this case, Eq. \ref{genint} reduces to the EW equation,
Eq. \ref{ew}. The resulting linear growth problem was also 
studied by Hammersley in a different context \cite{ham}. The values of
the critical exponents for this class are \cite{ew} 
\begin{equation}
\beta=1/4, ~z=2 ~~(\mathrm{EW ~~in}~ 1\mathrm{d}).
\label{ewexp}
\end{equation}

The Edwards-Wilkinson (EW) fixed point also controls the behavior in
systems where only odd-order terms $\lambda_{m}$ with $m \ge 3$ are present. The lowest-order nonlinearity is provided by the cubic term. Under
a scaling transformation $x \rightarrow x'=bx, t \rightarrow t'=b^{z}t,
h \rightarrow h'=b^{\alpha}h$ , we see 
that the cubic nonlinear term is marginal around the EW fixed point. 
Calculations based on mode coupling theory \cite{derrida2} and RG
\cite{MB4,paczuski} show that it is actually marginally irrelevant, 
leading to multiplicative logarithmic corrections to the power law: 
$S(0,t) \sim t^{1/2}(\ln t)^{1/4}$, a behavior that has been verified
by numerical simulation study \cite{paczuski,MB4,lebowitz}.

Even-order coefficients lead to a breaking of $h \rightarrow -h$
symmetry of Eq. \ref{genint}, causing a change in the universality
class. The second-order perturbation $\lambda_{2}$ is a relevant
perturbation for the EW fixed point that drives the system away from
the EW to a new KPZ fixed point. The KPZ fixed point is characterized
by the exponent values \cite{kpz} 
\begin{equation}
\beta=1/3, ~z=3/2 ~~(\mathrm{KPZ ~~in}~ 1\mathrm{d}).
\label{kpzexp}
\end{equation}

\begin{figure}[here]
\includegraphics[scale=0.4]{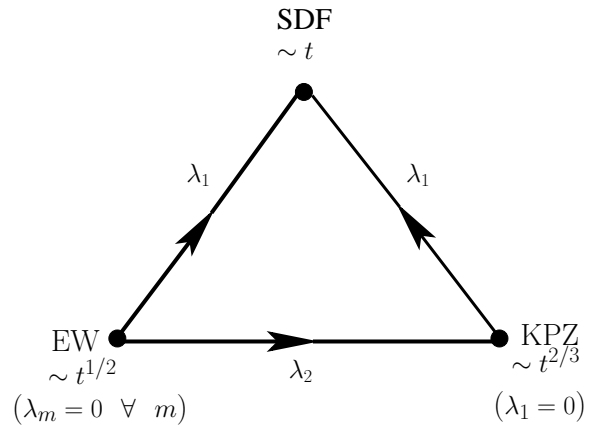}
\caption{Schematic representation of relative stabilities of
various fixed points, showing the associated behavior of the
variance $\sigma^{2}(L,t)$ of the displacement of the tagged particle in an infinite system.}
\label{flowdi}
\end{figure}

If $\lambda_{1} \ne 0$, under the scaling transformation $x
\rightarrow x'=bx, t \rightarrow t'=b^{z}t, h \rightarrow 
h'=b^{\alpha}h$,  both the EW and KPZ fixed points are unstable, 
and a flow towards a third fixed point (SDF) is generated (Fig.
\ref{flowdi}), as we show in
this paper. 

\section{Representation for an infinite system}
\label{repinfsys}
 On the basis of the discussion in Section \ref{themodel}, we 
 conclude that in the stationary state, the displacement of the 
 $n$-th tagged particle, $y(n,t)-y(n,0)$, has contributions from 
 three distinct physical sources, namely, systematic drift, sliding
 density fluctuations, and their dissipation. Correspondingly we write 
\begin{equation}
y(n,t)-y(n,0)\approx v_{P}t+t^{\alpha}G_{n}(t)+t^{\beta}\chi_{n}(t),
\label{post}
\end{equation}
where $\alpha=\alpha_{KPZ}=1/2$ is the roughness exponent while
$\beta=\beta_{KPZ}=1/3$ is
the growth exponent of the interface
equivalent to the ASEP. The representation, Eq. \ref{post}, is based on
the following arguments. First, the
coarse-grained density fluctuations for the ASEP are described by the KPZ
interface whose fluctuation profile moves with velocity $v_{K}-v_{P}$.
As a result, in time $t$, the particle senses the
fluctuations over a stretch of length $(v_{K}-v_{P})t$. From the scaling analysis of the interface summarized in Section
\ref{themodel}, the typical fluctuation over this distance scales as $[(v_{K}-v_{P})t]^{\alpha}\sim t^{\alpha}$. This fact is encoded in the second term in Eq.\ref{post}
where $G_{n}(t)$ is a random variable which depends only on the initial configuration, drawn from the
stationary ensemble, but is independent of the stochastic noise in
the evolution of configurations. The fluctuations arising from the
dissipation typically grow with time as $t^{\beta}$ and are represented by the
last term in Eq. \ref{post}. Here $\chi_{n}(t)$ is a random variable that depends only on the noisy 
history in the evolution of configurations, but is independent of the
initial configuration. 

We show below how the form, Eq. \ref{post}, accounts for the 
observed behavior of correlation functions for an infinite system.
\begin{itemize}
\item{
On averaging with respect to both the initial stationary ensemble
and stochastic evolution, we get
\begin{equation}
\overline{\langle y(n,t)-y(n,0) \rangle}\approx v_{P}t+t^{\alpha}\overline{G_{n}}
+t^{\beta}\langle \chi_{n}\rangle,
\end{equation}
where, as previously, the angular brackets denote averaging with
respect to the stochastic evolution, while the overbar is used to 
denote averaging with respect to the initial stationary ensemble.

It follows that for the variable defined in Eq. \ref{Deltadef}, 
\begin{equation}
\Delta_{n}(t)\approx t^{\alpha}(G_{n}-\overline{G_{n}})+t^{\beta}(\chi_{n}-
\langle \chi_{n} \rangle ).
\end{equation}
With $\alpha=1/2$ and $\beta=1/3$ for the ASEP, the first
term on the rhs dominates for large $t$. We 
recover the result for $\sigma^{2}(L,t)=\overline{\langle \Delta_{n}^{2}(t) \rangle}$ for an infinite system \cite{De Masi} 
(described in the Introduction), given by  
\begin{equation}
\lim_{L\rightarrow \infty} \sigma^{2}(L,t) \stackrel{t \rightarrow
\infty} \sim t \overline {(G_{n}-\overline {G_{n}})^{2}} \sim t.
\end{equation}
}
\item{ 
For fixed initial condition, averaging over noise history, we get
\begin{equation}
\langle y(n,t)-y(n,0) \rangle \approx v_{P}t+t^{\alpha}G_{n}+t^{\beta}
\langle\chi_{n}\rangle.
\end{equation}
Thus, for the displacement variable defined in Eq. \ref{ddef},
we have
\begin{equation}
d_{n}(t)\approx t^{\beta}[\chi_{n}-\langle \chi_{n}\rangle].
\end{equation}
The fluctuations in $d_{n}(t)$ are measured by $s^{2}(L,t)$ according
to the definition, Eq.
\ref{sdef}. Thus, one has
\begin{equation}
s^{2}(t)=\lim_{L \rightarrow \infty}s^{2}(L,t)\stackrel
{t \rightarrow \infty}\sim  t^{2\beta} \langle (\chi_{n}-\langle
\chi_{n} \rangle )^{2} \rangle \sim t^{2\beta}.
\end{equation}
With $\beta=1/3$, we get the result $s^{2}(t) \sim t^{2/3}$
\cite{vanb1} for the growth of the variance of the displacement of a
tagged particle in an infinite system, starting
from an arbitrary fixed initial configuration drawn from the
stationary ensemble, as discussed in the Introduction.
}
\end{itemize}

\section{Variances $\sigma^{2}(L,t)$ and
$s^{2}(L,t)$: Physical Arguments}
\label{physicalarg}
\subsection{Infinite system}
The variance $\sigma^{2}(t)$ of the displacement of a
tagged particle in an infinite system, averaged with respect to both the initial
stationary ensemble and stochastic evolution, behaves very differently
from the variance $s^{2}(t)$ where the averaging is with respect to only the stochastic
evolution. Such differences for an infinite system have also been
observed for shock fluctuations in the ASEP \cite{alexander}. In this subsection, we provide physical interpretations of the observed linear growth in time
for $\sigma^{2}(t)$ and the $t^{2/3}$ growth of $s^{2}(t)$.

\subsubsection{$\sigma^{2}(t)$}
 As the averaging in Eq. \ref{sigmadef} is with respect to both an initial
stationary ensemble and stochastic noise in the evolution of 
configurations, the average drift in time $t$ is $v_{P}t$, and fluctuations in the tagged particle displacement are
defined with respect to this. In the rest frame of the density
fluctuations, the tagged particle has an average velocity 
$u=v_{P}-v_{K}$. In time $t$, it traverses a sequence of density
fluctuations over a distance $(v_{P}-v_{K})t$, with each fluctuation adding a stochastic noise to the motion of the tagged
particle. The noise is uncorrelated since the stationary state configurations have a product measure. 
The variance $\sigma^{2}(L,t)$ is thus proportional to
$t$ for large $t$, by virtue of the central limit theorem. The
coefficient of proportionality $D_{0}$ is known to be $v_{P}=(p-q)(1-\rho)$
\cite{De Masi}. This expression for $D_{0}$ can be easily derived using the
above picture of drift of the tagged particles relative to the density
fluctuations \cite{MB2}.

\subsubsection{$s^{2}(t)$}
For $s^{2}(t)$, the fluctuations are measured by starting 
from a single fixed configuration, drawn from the ensemble of
stationary states. Thus, in every measurement, 
the particle passes through the \textit{same} sequence of density
fluctuations. However, dissipation of the density profile is different
from history to history.

\begin{figure}[here]
\begin{center}
\includegraphics[scale=0.7]{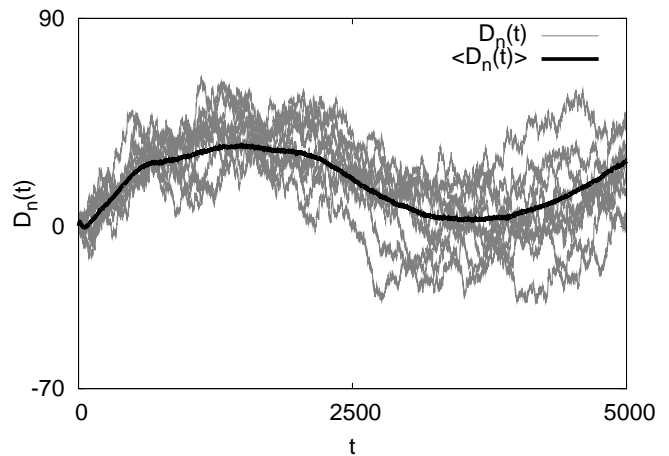}
\caption{The gray curves show the distance covered,
$D_{n}(t)$, in time $t$ about the mean $v_{P}t$ by the $n$-th 
tagged particle for $10$ Monte Carlo runs for a single fixed initial configuration,
drawn from the stationary ensemble of the TASEP. The black curve shows the mean 
displacement, $\langle D_{n}(t) \rangle$, obtained by averaging 
over $500$ histories. Here, $L=10,000$, the particle density $\rho=0.25$.}
\label{10000-2500-stavst}
\end{center}
\end{figure}

Define the deviation in the displacement of a tagged particle 
from $v_{P}t$ by
\begin{equation}
D_{n}(t)=y(n,t)-y(n,0)-v_{P}t.
\end{equation}
The dissipation in the density profile being different for different histories shows up in
$D_{n}(t)$, as depicted by the
gray trajectories in Fig. \ref{10000-2500-stavst}.
 Notice that the mean over histories, $\langle D_{n}(t) \rangle$, is
 nonvanishing, as depicted by the black curve in Fig. \ref{10000-2500-stavst}. 

Since the variance $s^{2}(t)$ of the fluctuation in Eq. \ref{sdef} is measured with
respect to $v_{P}t+\langle D_{n}(t) \rangle$, it is able to sense the
dissipation of the density profile in time, without the effects of SDF.
From Eq. \ref{post}, with $\alpha=1/2$ and $\beta=1/3$, we get $D_{n}(t)=t^{1/2}G_{n}(t)+t^{1/3}\chi_{n}(t)$;
on averaging over stochastic evolution, we get
\begin{equation}
\langle D_{n}(t) \rangle=t^{1/2}G_{n}(t)+t^{1/3}\langle \chi_{n}(t)\rangle.
\label{avdnt}
\end{equation}
The first term in Eq. \ref{avdnt} dominates, and thus at large $t$,
$\langle D_{n}(t) \rangle$ is primarily determined by the pattern
of spatial density fluctuations in the initial configuration. Ignoring
the dissipative component (the second term in Eq. \ref{avdnt}), we
expect $\langle D_{n}(t) \rangle$ to be given by $D_{n}^{*}(t) \approx
(p-q)\int_{0}^{t}dt'\{(1-\rho(x))-(1-\rho)\}$. Making a change of
variable from $t$ to $x$, and noting that the Jacobian of transformation
is $1/(v_{P}-v_{K})$, we get
\begin{equation}
D_{n}^{*}(t) \approx \frac{(p-q)}{v_{P}-v_{K}}\int_{y(n,0)}^{y(n,0)+(v_{P}-v_{K})t}dx\{(1-\rho(x))-(1-\rho)\}.
\label{denprof}
\end{equation}
The upper limit of the integral incorporates the relative
distance moved by the particle, $(v_{P}-v_{K})t$, neglecting fluctuations of
$O(t^{1/2})$ coming from the second term in Eq. \ref{post}. 

\begin{figure}[!h]
\begin{center}
\includegraphics[angle=270,scale=0.3]{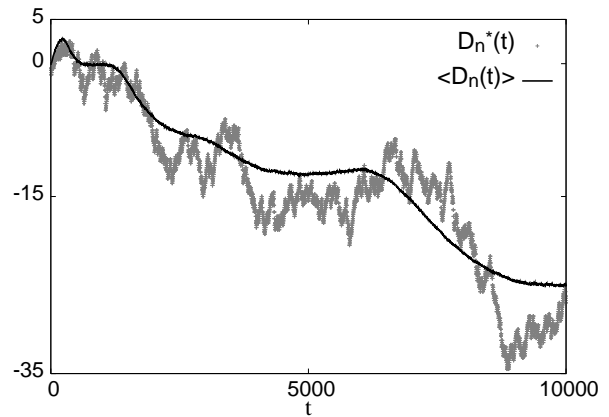}
\caption{Comparison between $\langle D_{n}(t)
\rangle$ and $D_{n}^{*}(t)$ for a fixed initial stationary 
configuration of the TASEP.  $\langle D_{n}(t) \rangle$ was
obtained from simulations, and involves averaging over $1000$ histories.
$D_{n}^{*}(t)$ was obtained by integrating the initial density profile
according to Eq. \ref{denprof} (see text). Here, $L=10,000$, while the density $\rho=0.25$.}
\label{10000-2500-confg-9fig}
\end{center}
\end{figure}

In Fig. \ref{10000-2500-confg-9fig}, we have compared $\langle
D_{n}(t) \rangle$ with $D_{n}^{*}(t)$ for a randomly chosen initial configuration. $\langle D_{n}(t) \rangle$ was
obtained from simulations, while $D_{n}^{*}(t)$ was obtained by integrating the 
initial density profile, following Eq. \ref{denprof}. We see that there
is good agreement between the signals for $\langle
D_{n}(t) \rangle$ and $D_{n}^{*}(t)$. Fluctuations over a distance $\delta x$ dissipate in a
time $t \sim (\delta x)^{z}$ with $z=3/2$. Thus, with increasing $t$,
spatial fluctuations in the initial profile are smeared out over larger
distances, a fact which is borne out by Fig. \ref{10000-2500-confg-9fig}. We have also measured the overlap function for the signs of the
two signals,  $\langle D_{n}(t) \rangle$ and $D_{n}^{*}(t)$, 
for $10$ different randomly chosen initial configurations at 
stationarity, and found a mean value $0.8$, which indicates a fairly
good degree of correlation between the two signals.

\subsection{Finite system}
We now turn to the behavior of $\sigma^{2}(L,t)$ and $s^{2}(L,t)$ on a
finite lattice, and explain the differences in their behavior on the basis of SDF.
\subsubsection{$\sigma^{2}(L,t)$}
 In the rest frame of the density
fluctuations, a tagged particle has velocity $u=v_{P}-v_{K}$ and would
take time $T=L/u$ to return to its initial environment of density
fluctuations. Thus the variance of the tagged particle displacement should increase from $0$ at $t=0$, reach a maximum at around
$T/2$, and come down to almost zero at $t=T$. The difference from zero
at $t=T$ is due to dissipation in the density profile arising from stochasticity in the 
dynamics. This scenario recurs at integral $T$'s; the time period of
oscillations is $T=L/u=L/[(p-q)\rho]$ 

The oscillations do not continue forever because of the cumulative
effect of dissipation. An
estimate for the time taken to completely dissipate an initial 
pattern of density fluctuations is $T_{2} \sim L^{z}$, where the
dynamical exponent $z$ is known to have the value $3/2$ \cite{gwa}.
For times $t \gg T_{2}$, when the initial density profile has completely
dissipated away, the fluctuations are entirely due to the motion of the
center-of-mass: $\sigma^{2}(L,t) \sim D(L)t$ with $D(L) \sim \frac{1}{\sqrt{L}}$. The
scaling of $D(L)$ with the system size $L$ is obtained by matching the
behavior of $\sigma^{2}(L,t)$ across $T_{2}$.

\begin{figure}[h!]
\begin{center}
\includegraphics[scale=0.7]{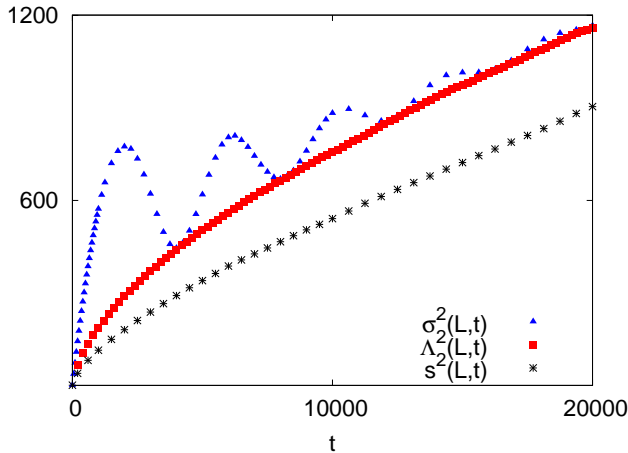}
\caption{(Color online) 
MC simulation results for various tagged particle correlations. The
topmost curve refers to the variance $\sigma^{2}(L,t)$ of the
displacement when the averaging is over both the initial stationary ensemble and
stochastic evolution. The middle curve is the sliding tag correlation
function $\Lambda^{2}(L,t)$ defined in the text; it coincides with
$\sigma^{2}(L,t)$ at the local minima of the latter. The lowermost curve
shows the variance $s^{2}(L,t)$ when the averaging is only over
stochastic evolution.
Here, $\rho=0.25$. The system size is $1024$. The averaging is
over $10^{5}$ MC runs.}
\label{slidingtag}
\end{center}
\end{figure}

The lower envelope $\Lambda^{2}(L,t)$ of the oscillations in
$\sigma^{2}(L,t)$ is determined by the
dissipation of the stationary state density profile, and can be studied
by the method of sliding tags \cite{MB1,MB2,MB3,MB4,paczuski,MB6,ddas,alcaraz}. In order to monitor the dissipation of the moving density profile, we need to correlate, at
different times, the location of two different particles. Since in the frame of the density fluctuations, particles move with velocity 
$u=v_{P}-v_{K}$, we need to examine the function 
\begin{equation}
\Lambda^{2}(L,t)=\overline{\langle
\left[y(n',t)-y(n,0)\right]^{2}\rangle}-\overline{\langle \left[
y(n',t)-y(n,0) \right] \rangle}~^{2}, 
\end{equation}
where 
\begin{equation}
n'=n-\rho ut.
\label{slidetag}
\end{equation}
The tag shift $\rho ut$ accounts for the relative motion of the
particles and the density profile, and ensures that the time evolution
of the same density patch is being recorded at every instant.
Equation \ref{slidetag}, representing sliding of the  
tag, is tantamount to a Galilean shift that gets rid of the drift
term, $u\frac{\partial h}{\partial x}$, in Eq. \ref{kpzdrift}. In Fig \ref{slidingtag}, we show how the 
method of sliding tags discussed above gives the lower envelope $\Lambda^{2}(L,t)$ of the oscillatory quantity 
$\sigma^{2}(L,t)$. This envelope grows with time 
as $t^{2/3}$ until times $\sim L^{3/2}$, beyond which $\Lambda^{2}(L,t) \sim D(L)t$ with $D(L) \sim \frac{1}{\sqrt{L}}$. Thus,
beyond $T_{2} \sim L^{3/2}$, there is no distinction between the
temporal behavior of $\sigma^{2}(L,t)$ and $\Lambda^{2}(L,t)$. 
\subsubsection{$s^{2}(L,t)$}
Since we always use the same initial condition in this case, the particle moves through the same sequence of density
fluctuations in every measurement.  
Nevertheless, the dissipation of the density profile is different for
different histories, and $s^{2}(L,t)$ captures this. Typical fluctuation
grows with time as $t^{1/3}$, following Eq. \ref{post}. This leads to $s^{2}(L,t) \sim t^{2/3}$. This behavior continues until $t \gg T^{*} \sim L^{3/2}$ when
the fluctuations are due to the diffusive motion of the center-of-mass. This gives $s^{2}(L,t) \sim D(L)t$ with $D(L)
\sim \frac{1}{\sqrt{L}}$. One obtains the scaling of $D(L)$ with the
system size $L$ by a simple match in the behavior of $s^{2}(L,t)$ across $T^{*}\sim L^{3/2}$.

\section{Linear Interface Model: Analytical Results}
\label{exactsol}
The existence of size-dependent time scales in $\sigma^{2}(L,t)$ and $s^{2}(L,t)$ for the ASEP,
discussed in this paper, can be explained qualitatively by exactly solving the continuum equation, Eq. 
\ref{kpzdrift}, in the limit $\lambda \rightarrow 0$. This is
possible as, in this limit, Eq. \ref{kpzdrift} is linear, and 
hence, solvable. We refer to the corresponding interface as a 
linear interface. The relevant time evolution equation is
\begin{equation}
\frac{\partial h(x,t)}{ \partial t}=v_{P}+\Gamma \frac
{\partial^{2}h}{\partial x^{2}}+u\frac{\partial h}{\partial x}+\eta(x,t).
\label{lin1}
\end{equation}
In this section, we outline the exact computation of the two quantities
$\sigma^{2}(L,t)$ and $s^{2}(L,t)$ for the linear 
interface. We will see that the drift term, $u\frac{\partial h}
{\partial x}$ in Eq. \ref{lin1}, that makes it different from the 
usual EW equation, plays a crucial role in determining the temporal
behavior of these functions.

Our aim is to compute $\sigma^{2}(L,t)$ and $s^{2}(L,t)$ for
the linear interface. For this, we look at the function $C_{L}(t_{0},t_{0}+t)$,
defined in Eq. \ref{cldef}, which, utilizing the mapping outlined in
Table \ref{mapping} with $n$ replaced by the continuous variable $x$,
reads
\begin{widetext}
\begin{equation}
C_{L}(t_{0},t_{0}+t)=\langle[h(x,t_{0}+t)-h(x,t_{0})-
\langle[h(x,t_{0}+t)-h(x,t_{0})]\rangle]^{2}\rangle.
\label{clhtdef}
\end{equation}
\end{widetext}
As discussed in Section \ref{themodel}, $\sigma^{2}(L,t)$ can be obtained
from the limiting behavior of the function $C_{L}(t_{0},t_{0}+t)$ in the limit $t_{0}
\rightarrow \infty$. On the other hand, one recovers $s^{2}(L,t)$ 
from the limiting behavior of the function $C_{L}(t_{0},t_{0}+t)$ in the limit $t_{0}
\rightarrow 0$.

To compute $C_{L}(t_{0},t_{0}+t)$, we need to solve Eq. \ref{lin1} for $h(x,t)$.
Before doing so, we note that the constant term on the right of Eq. \ref{lin1} can be gotten
rid of by going to a co-moving frame moving with velocity $v_{P}$. This is equivalent to making the transformation $h \rightarrow
h+v_{P}t$. The resultant equation is the usual time-evolution equation
for the EW interface, Eq. \ref{ew}, with an additional drift term,
$u\frac{\partial h}{\partial x}$, and can be solved by going to
Fourier space. To this end, utilizing the boundary condition $h(x \pm
L,t)=h(x,t)$ on the height variable, we write $h(x,t)$ in terms of its
Fourier modes $\tilde{h}(m,t)$. We have 
\begin{equation}
h(x,t)=\sum\limits_{m=-\infty}^{\infty}\tilde{h}(m,t)
e^{\frac{i2\pi m}{L}(x+ut)}.
\end{equation}
Thus, $\tilde{h}(m,t)=\frac{1}{L}\int_{0}^{L} dx~ h(x,t)
e^{-\frac{i2\pi m}{L}(x+ut)}$. Equation \ref{lin1} now reads
\begin{equation}
\frac{\partial \tilde{h}(m,t)}{ \partial t}=-\Gamma 
\frac{4\pi^{2}m^{2}}{L^{2}}\tilde{h}(m,t)+\tilde{\eta}(m,t),
\label{foureq}
\end{equation}  
where $\tilde{\eta}(m,t)=\frac{1}{L}\int_{0}^{L}dx~ 
\eta(x,t)e^{-\frac{i2\pi m}{L}(x+ut)}$. Utilizing $\langle
\eta(x,t)\eta(x',t')\rangle=2A
\delta(x-x')\delta(t-t')$, we have, for the Fourier modes,
\begin{equation}
\langle \tilde{\eta}(m,t)\tilde{\eta}(m',t')\rangle=\frac{2A}{L}
\Delta_{m,-m'}\delta(t-t'),
\label{etacorr}
\end{equation}
where $\Delta_{m,n}$ is the Kronecker delta. 

Equation \ref{foureq} can be solved for $\tilde{h}(m,t)$ to get
\begin{eqnarray}
\tilde{h}(m,t)&=&\tilde{h}(m,0)e^{-\Gamma \frac{4\pi^{2}m^{2}}
{L^{2}}t}\nonumber \\ &+&e^{-\Gamma \frac{4\pi^{2}m^{2}}{L^{2}}t}
\int_{0}^{t} dt' ~\tilde{\eta}(m,t')e^{\Gamma \frac{4\pi^{2}m^{2}}
{L^{2}}t'}. \nonumber \\
\label{htfour}
\end{eqnarray}
The condition $\langle \eta(x,t) \rangle=0$ gives, for its 
Fourier modes, $\langle \tilde{\eta}(m,t) \rangle=0 ~\forall ~m$.
Using this in Eq. \ref{htfour}, we get $\langle \tilde{h}(m,t)\rangle
=\langle \tilde{h}(m,0) \rangle
e^{-\Gamma \frac{4\pi^{2}m^{2}}{L^{2}}t}$.  
 
From Eq. \ref{htfour}, with the help of Eq. \ref{etacorr}, we get
\begin{eqnarray}
&&\langle \tilde{h}(m,t)\tilde{h}(m',t') \rangle= \langle \tilde{h}(m,0)\tilde{h}(m',0)\rangle e^{-\Gamma
\frac{4\pi^{2}}{L^{2}}(m^{2}t+m'^{2}t')} \nonumber \\
&+&\frac{AL}{4\pi^{2}\Gamma}\frac{\Delta_{m,-m'}}{m^{2}}
\left[e^{-\Gamma \frac{4\pi^{2}m^{2}}{L^{2}}|t-t'|}-
e^{-\Gamma \frac{4\pi^{2}m^{2}}{L^{2}} (t+t')}\right].\nonumber \\
\label{hthtfour}
\end{eqnarray}
\subsection{$\sigma^{2}(L,t)$}
Here, we need to take the limit $t_{0} \rightarrow \infty$ in the
function $C_{L}(t_{0},t_{0}+t)$. Thus, 
we will compute
\begin{widetext}
\begin{equation}
\sigma^{2}(L,t)=\lim_{t_{0} \rightarrow \infty}C_{L}(t_{0},t_{0}+t)=\lim_{t_{0} \rightarrow \infty}\langle[h(x,t_{0}+t)-h(x,t_{0})-\langle[h(x,t_{0}+t)-h(x,t_{0})]\rangle]^{2}\rangle.
\label{clinft}
\end{equation}
\end{widetext}
In the limit $t_{0} \rightarrow \infty$, the quantity $\langle h(x,t_{0}+t) -
h(x,t_{0})\rangle$ goes to zero. Thus, the
function $\sigma^{2}(L,t)$ reduces to $\sigma^{2}(L,t)=\lim_{t_{0}\rightarrow \infty}\langle [h(x,t_{0}+t)-h(x,t_{0})]^{2}\rangle$.  

Utilizing Eq. \ref{hthtfour}, we get
\begin{eqnarray}
\lefteqn{\langle h(x,t) h(x,t') \rangle=
\langle\tilde{h}^{2}(0,0)\rangle+\frac{2A}{L}\mathrm{min}(t,t')}\nonumber\\
&&+\frac{AL}{4\pi^{2}\Gamma}\sum\limits_{m=-\infty,m \ne 0}^{\infty}
\frac{1}{m^{2}}\left[e^{-\Gamma\frac{4\pi^{2}m^{2}}{L^{2}}|t-t'|}-
e^{-\Gamma\frac{4\pi^{2}m^{2}}{L^{2}}(t+t')}\right]\nonumber \\
&& ~~~~~~~~~~~~~~~~~~~~~~~~~~~~~~~~~~~~~~~~~~~~\times ~~
e^{\frac{i2\pi mu}{L}(t-t')}.\nonumber \\
\label{hthtx}
\end{eqnarray}
In the last equation, the contribution from the initial condition,
$\langle \tilde{h}(m,0)\tilde{h}(m',0)\rangle e^{-\Gamma 
\frac{4\pi^{2}m^{2}}{L^{2}}(m^{2}t+m'^{2}t')}$ for $m,m' \ne 0$ has been dropped,
since, eventually when we set $t=t_{0}$ and $t'=t_{0}+t$ and let 
$t_{0} \rightarrow \infty$, this exponential term goes to $0$.
 By combining the $m$ and $-m$ terms in the summation in Eq. \ref{hthtx}, we get
\begin{eqnarray}
\langle h(x,t)h(x,t')\rangle=\langle\tilde{h}^{2}(0,0)\rangle+\frac{2A}{L}\mathrm{min}(t,t')\nonumber \\
+\frac{AL}{2\pi^{2}\Gamma}\sum\limits_{m=1}^{\infty}\frac{1}
{m^{2}}\left[e^{-\Gamma \frac{4\pi^{2}m^{2}}{L^{2}}|t-t'|}-
e^{-\Gamma\frac{4\pi^{2}m^{2}}{L^{2}}(t+t')}\right]\nonumber \\
\times~~ \cos\left(\frac{2\pi m u}{L}(t-t')\right). \nonumber \\
\label{hthtx2}
\end{eqnarray}
Utilizing this expression in $C_{L}(t_{0}+t,t_{0})$, and then
taking the limit $t_{0} \rightarrow \infty$, we finally get
\begin{equation}
\sigma^{2}(L,t)=\frac{2A}{L}t+\frac{AL}{\pi^{2}\Gamma}
\sum_{m=1}^{\infty}\frac{1}{m^{2}}\left[1-e^{-\Gamma \frac
{4\pi^{2}m^{2}}{L^{2}}t}\cos\left(\frac{2\pi mu}{L}t\right)\right].
\label{sigmatlfin}
\end{equation}
This is an exact formula for $\sigma^{2}(L,t)$ within the 
linear model. Next, we consider the various limits.
\begin{itemize}
\item{
$t \ll L/u$. Let $\frac{2\pi ut}{L}m=k$. With $\Delta m=1$, we get 
$\Delta k=\frac{2\pi ut}{L}$. Now, $ut \ll L$ implies that $\Delta k$ is small,
and hence, one can replace the sum over $m$ in Eq. \ref{sigmatlfin} 
by an integral over $k$ to get
\begin{equation}
\sigma^{2}(L,t)\approx \frac{2Aut}{\pi \Gamma}
\int_{0}^{\infty}\frac{dk}{k^{2}}\left[1-e^{-\frac{\Gamma k^{2}}
{u^{2}t}}\cos(k)\right],
\end{equation}
where, in obtaining the last equation, we have dropped the first term on the right of Eq.
\ref{sigmatlfin} in comparison with the second term. 
The integral on the right can be done exactly, see Appendix \ref{integralev}. 
Using its value, we get
\begin{equation}
\sigma^{2}(L,t)=\frac{Au}{\Gamma}\left[2\frac{\sqrt{\Gamma}}
{\sqrt{\pi}u}\sqrt{t}e^{-\frac{u^{2}t}{4\Gamma}}+t ~\mathrm{erf}
\left(\frac{u}{2\sqrt{\Gamma}}\sqrt{t}\right)\right].
\end{equation}
Now we consider two cases:

(a) $t \ll \frac{4\Gamma}{u^{2}}$: In this limit, the error function is
approximately zero, and we get
\begin{equation}
\sigma^{2}(L,t) \approx \frac{2A}{\sqrt{\pi \Gamma}}\sqrt{t}.
\end{equation}

(b) $\frac{4\Gamma}{u^{2}} \ll t \ll \frac{L}{u}$. Here, 
\begin{equation}
\sigma^{2}(L,t) \approx \frac{Au}{\Gamma}t.
\end{equation}
}
\item{
$ t \sim \frac{L}{u}$: Let $t=\frac{nL}{u}$ with $n \in I$. Also,
let $\frac{2\pi m\sqrt {\Gamma}}{\sqrt{uL}}=k$. Substituting in Eq.
\ref{sigmatlfin} after replacing, for large $L$, the sum over $m$ by
an integral over $k$, we get
\begin{equation}
\sigma^{2}(L,t=\frac{nL}{u})\approx \frac{2At}{L}+\frac{2A}
{\pi}\sqrt{\frac{L}{u\Gamma}}\int_{0}^{\infty}\frac{dk}{k^{2}}(1-e^{-k^{2}n}). 
\end{equation}
The integral on the right can be done exactly, and its value is 
$\sqrt{n\pi}$. This gives
\begin{equation}
\sigma^{2}(L,t=\frac{nL}{u})\approx \frac{2A}{L}t+\frac{2A}
{\sqrt{\pi \Gamma}}\sqrt{t}.
\label{sigmatl2}
\end{equation}

For large $L$, keeping $t$ fixed, when the first term on the 
rhs goes to zero, we have
\begin{equation}
\sigma^{2}(L,t=\frac{nL}{u})\approx \frac{2A}{\sqrt{\pi \Gamma}}
\sqrt{t}.
\end{equation}
}
\item{
$t \gg L^{2}$: In this limit, the exponential term in the sum on the 
rhs of Eq. \ref{sigmatlfin} drops out for all $m$ to give
\begin{eqnarray}
\sigma^{2}(L,t)\approx \frac{2A}{L}t+\frac{AL}{\pi^{2}\Gamma}
\sum\limits_{m=1}^{\infty}\frac{1}{m^{2}}\nonumber \\=\frac{2A}{L}t+
\frac{AL}{\pi^{2}\Gamma}\left(\frac{\pi^{2}}{6}\right).
\end{eqnarray}
In the limit of large $t$, the first term on the rhs dominates so
that
\begin{equation}
\sigma^{2}(L,t)\approx\frac{2A}{L}t.
\end{equation}
}
Note that the large time behavior of $\sigma^{2}(L,t)$ is determined by
the zero mode ($m=0$) in the Fourier expansion of $h(x,t)$.

With the values of the exponents $\beta$ and $z$ for EW given in
Eq. \ref{ewexp}, we summarize the behavior of $\sigma^{2}(L,t)$ for
the linear interface in Table \ref{sigmatable}.
\begin{table}[h!]
\begin{tabular}{|c|c|}\hline
$t \ll T_{1} \sim L$: & \\
(a) $t \ll \frac{4\Gamma}{u^{2}}$. & $\sigma^{2}(L,t) \approx
\frac{2A}{\sqrt{\pi \Gamma}}\sqrt{t}$. \\
(b) $\frac{4\Gamma}{u^{2}} \ll t \ll T_{1}$. & $\sigma^{2}(L,t) 
\approx \frac{Au}{\Gamma}t$. \\ \hline
$T_{1} \ll t \ll T_{2}\sim L^{z}. $&$\sigma^{2}(L,t)\approx \frac{2A}{\sqrt{\pi 
\Gamma}}{t}^{2\beta}.$ \\
$z=z_{EW}=2$.& $\beta=\beta_{EW}=1/4$.\\
$t=\frac{nL}{u}$ with $n \in I$. &  \\ \hline
$t \gg T_{2}$.&$\sigma^{2}(L,t) \approx \frac{2A}{L}t$. \\
& The diffusion constant $D(L)=\frac{2A}{L}$. \\ \hline
\end{tabular}
\caption{Behavior of $\sigma^{2}(L,t)$ in different time regimes for the linear interface.}
\label{sigmatable}
\end{table}

The scaling of the diffusion constant $D(L)$ as the inverse of the
system size in the regime $t \gg T_{2} \sim L^{z}$ with $z=z_{EW}=2$ can be obtained by matching the behavior of the $\sigma^{2}(L,t)$ 
across the time $T_{2}$. 

Let $\sigma^{2}(L,t) \sim D(L)t$ with $D(L) \sim L^{\gamma}$ for $t \gg T_{2}$. 

On the lower side of $T_{2}$, we have $\sigma^{2}(L,t) \sim t^{2\beta}$. 

Then, to match the $t$ and $L$ dependence at $T_{2}$, we must have
\begin{equation}
(T_{2})^{2\beta} \sim D(L)T_{2}.
\end{equation}

This gives $2\beta=\gamma/z+1$ whence, with $z=2$ and $\beta=1/4$ for the 
linear model, $\gamma=-1$.

On the basis of the above results for the linear model, we 
expect the time-dependence of $\sigma^{2}(L,t)$ for the
KPZ class as in Table \ref{sigmatablekpz}.
\begin{table}[h!]
\begin{tabular}{|c|c|}\hline
$t \ll T_{1} \sim L$: & \\
(a) $t \ll \frac{4\Gamma}{u^{2}}$. & $\sigma^{2}(L,t) \sim \sqrt{t}$. 
\\
(b) $\frac{4\Gamma}{u^{2}} \ll t \ll T_{1}$. & $\sigma^{2}(L,t) \sim
t$. \\ \hline
$T_{1} \ll t \ll T_{2}\sim L^{z}.$&$\sigma^{2}(L,t)\sim {t}^{2\beta}$. \\
$z=z_{KPZ}=3/2$. &$\beta=\beta_{KPZ}=1/3$.  \\
$t=\frac{nL}{u}$ with $n \in I$. & \\ \hline 
$t \gg T_{2}$. & $\sigma^{2}(L,t) \sim \frac{1}
{\sqrt{L}}t$. \\
 & The diffusion constant $D(L) \sim 1/\sqrt{L}$. \\ 
\hline
\end{tabular}
\caption{Behavior of $\sigma^{2}(L,t)$ in different time regimes for the
KPZ interface.}
\label{sigmatablekpz}
\end{table}
Here, $z=z_{KPZ}=3/2$ and $\beta=\beta_{KPZ}=1/3$ so that $\gamma=-
1/2$. This value for $\gamma$ leads to the inverse square root scaling
of the diffusion constant $D(L)$ with system size $L$. 
\end{itemize}
\subsection{$s^{2}(L,t)$}
Here, we need to take $t_{0} \rightarrow 0$ in the
function $C_{L}(t_{0},t_{0}+t)$. Thus, we will compute
\begin{eqnarray}
s^{2}(L,t)&=&\lim_{t_{0} \rightarrow 0}C_{L}(t_{0},t_{0}+t)
\nonumber \\
&=&\langle[h(x,t)-h(x,0)-\langle[h(x,t)-h(x,0)]\rangle]^{2}\rangle. 
\nonumber \\
\label{cl0t}
\end{eqnarray}
Now, from Eq. \ref{htfour}, we have
\begin{eqnarray}
h(x,t)-h(x,0)=\sum\limits_{m=-\infty}^{\infty}[\tilde{h}(m,0)
e^{-\Gamma \frac{4\pi^{2}m^{2}}{L^{2}}t}+\nonumber \\
e^{-\Gamma\frac{4\pi^{2}m^{2}}{L^{2}}t}\int_{0}^{t}dt' 
\tilde{\eta}(m,t')e^{\Gamma \frac{4\pi^{2}m^{2}}{L^{2}}t'}] 
\times e^{\frac{i2\pi m}{L}(x+ut)}\nonumber \\ -
\sum\limits_{m=-\infty}^{\infty}\tilde{h}(m,0)e^{\frac{i2\pi m}{L}x}.
\nonumber \\
\end{eqnarray}
Noting that every time we start from the same initial condition so
that $\langle \tilde{h}(m,0)\rangle=\tilde{h}(m,0)$, we have
\begin{eqnarray}
h(x,t)-h(x,0)-\langle[h(x,t)-h(x,0)]\rangle =\nonumber \\
\sum\limits_{m=-\infty}^{\infty}e^{-\Gamma\frac{4\pi^{2}m^{2}}{L^{2}}}e^{\frac{i2\pi
m}{L}(x+ut)}
\int_{0}^{t}dt' \tilde{\eta}(m,t')e^{\Gamma \frac{4\pi^{2}m^{2}}{L^{2}}t'}. \nonumber \\
\label{cl0t2}
\end{eqnarray}
Utilizing Eq. \ref{cl0t2} and Eq. \ref{etacorr} in Eq. \ref{cl0t}, we get, after a few
steps,
\begin{equation}
s^{2}(L,t)=\frac{AL}{4\pi^{2}\Gamma}\sum\limits_{m=-\infty}^
{\infty}\frac{1}{m^{2}}(1-e^{-\Gamma \frac{8\pi^{2}m^{2}}{L^{2}}t}).
\end{equation}
Separating out the $m=0$ mode and exploiting the $m \rightarrow -m$ 
symmetry of the remaining terms, we get the final expression for
$s^{2}(L,t)$ as
\begin{equation}
s^{2}(L,t)=\frac{2A}{L}t+\frac{AL}{2\pi^{2}\Gamma}\sum\limits_
{m=1}^{\infty}\frac{1}{m^{2}}(1-e^{-\Gamma \frac{8\pi^{2}m^{2}}{L^{2}}t}).
\label{s2final}
\end{equation}
Next, we will consider the various limits.
\begin{itemize}
\item{
$t \ll L^{2}$: Let $k=\frac{2\sqrt{2\Gamma}\pi m}{L}\sqrt{t}$. Since
$\Delta m=1$, we have $\Delta k=\frac{2\sqrt{2\Gamma}\pi}{L}\sqrt{t}$ is 
small for $t \ll L^{2}$. This allows the sum over $m$ in Eq.
\ref{s2final} to be replaced by an integral over $k$ to give
\begin{equation}
s^{2}(L,t)=\frac{2A}{L}t+\frac{A}{\pi}\sqrt{\frac{2}{\Gamma}}
\sqrt{t}\int_{\frac{2\sqrt{2\Gamma}\pi\sqrt{t}}{L}}^{\infty}
\frac{dk}{k^{2}}(1-e^{-k^{2}}).
\label{s2tl1}
\end{equation}
Since $t \ll L^{2}$, the lower limit of the integral on the right
can be taken to be $0$. Then the integral can be done exactly, 
and its value is $\sqrt{\pi}$. For large $L$, keeping $t$ fixed, 
the first term on the right of Eq. \ref{s2tl1} goes to zero, and
we get
\begin{equation}
s^{2}(L,t) \approx A\sqrt{\frac{2}{\pi\Gamma}}\sqrt{t}. 
\end{equation}
}
\item{
$t \gg L^{2}$: In this limit, the exponential term in the sum on the
rhs of Eq. \ref{s2final} gets damped out for all $m$ so that we
finally have
\begin{eqnarray}
s^{2}(L,t)\approx \frac{2A}{L}t+\frac{AL}{2\pi^{2}
\Gamma}\sum\limits_{m=1}^{\infty}\frac{1}{m^{2}}\nonumber \\=
\frac{2A}{L}t+\frac{AL}{2\pi^{2}\Gamma}\left(\frac{\pi^{2}}{6}\right).
\end{eqnarray}
For large $t$, the first term on the rhs dominates so that
\begin{equation}
s^{2}(L,t)\approx\frac{2A}{L}t.
\end{equation}
}
Here, as for the function $\sigma^{2}(L,t)$, we see that the large time behavior of $s^{2}(L,t)$ is determined by
the zero mode ($m=0$) in the Fourier expansion of $h(x,t)$.
\end{itemize}
Knowing the values of the exponents $\beta$ and $z$ for EW, given
in Eq. \ref{ewexp}, we summarize the behavior of
$s^{2}(L,t)$ for the linear interface in Table \ref{stable}.
\begin{table}
\begin{tabular}{|c|c|}\hline
$t \ll T^{*} \sim L^{z}$. & $s^{2}(L,t) \approx 
Ct^{2\beta}$. \\ 
$z=z_{EW}=2$. & $\beta=\beta_{EW}=1/4$. \\
&$C=A\sqrt{\frac{2}{\pi\Gamma}}$. \\ \hline
$t \gg T^{*} $.& $s^{2}(L,t) \approx \frac{2A}{L}t$. \\
$z=z_{EW}=2$.& The diffusion constant $D(L)=\frac{2A}{L}$. \\ \hline
\end{tabular}
\caption{Behavior of $s^{2}(L,t)$ in different time regimes for the linear interface.}
\label{stable}
\end{table}
The fact that $D(L)$ scales as the inverse of the system size in the
regime $t \gg T^{*} \sim L^{z}$ with $z=z_{EW}=2$, can be obtained by matching the
behavior of $s^{2}(L,t)$ across the time $T^{*}$.

Thus, let $s^{2}(L,t) \sim D(L)t$ with $D(L) \sim L^{\gamma}$ for
$t \gg T^{*}$.

For $t \ll T^{*}$, we have $s^{2}(L,t) \sim t^{2\beta}$. 

Then, to match the $t$ and $L$ dependence at $T^{*}$, we must have
\begin{equation}
(T^{*})^{2\beta} \sim D(L)T^{*}.
\end{equation}

This gives $2\beta=\gamma/z+1$. With $\beta=1/4$ and $z=2$ for the
linear model, $\gamma=-1$.   

The above results lead us to expect the time-dependence
of $s^{2}(L,t)$ for the KPZ class as in Table \ref{stablekpz}.
\begin{table}[h!]
\begin{tabular}{|c|c|} \hline
$t \ll T^{*} \sim L^{z}$. & $s^{2}(L,t) \sim t^{2\beta}$. \\
$z=z_{KPZ}=3/2$. & $\beta=\beta_{KPZ}=1/3$. \\ \hline
$t \gg T^{*}$. & $s^{2}(L,t) \sim \frac{1}{\sqrt{L}}t$.
\\
& The diffusion constant $D(L) \sim 1/\sqrt{L}$. \\ 
\hline
\end{tabular}
\caption{Behavior of $s^{2}(L,t)$ in different time regimes for the
KPZ interface.}
\label{stablekpz}
\end{table}

Here, $\beta=\beta_{KPZ}=1/3$ and $z=z_{KPZ}=3/2$ so that 
$2\beta=\gamma/z+1$ gives $\gamma=-1/2$. This explains the 
inverse square root scaling of the diffusion constant $D(L)$ 
with system size $L$.  

Note that from Table \ref{stablekpz}, it follows that for an infinite
system when the time-scale $T^{*}$ diverges, one recovers the result observed by van Beijeren that 
$\lim_{L\rightarrow \infty}s^{2}(L,t)=s^{2}(t) 
\sim t^{2/3}$ \cite{vanb1}.

To conclude this section, we reiterate that the exact solution
for the linear interface qualitatively explains the occurrence of
characteristic oscillations and different $L-$dependent regimes in
the variance of the displacement on a finite lattice, as observed, for instance, in the the Monte Carlo simulations for the ASEP. We also note that in
the large time limit ($t \gg L^{z}$ with $z=3/2$ for the ASEP and $z=2$
for the linear interface), both $\sigma^{2}(L,t)$ and $s^{2}(L,t)$ grow
linearly in time with the same constant of proportionality or the
diffusion constant $D(L)$. For the linear interface, $D(L) \sim
\frac{1}{L}$, while $D(L) \sim \frac{1}{\sqrt{L}}$ for the ASEP.
\section{Correspondence of the linear model to different interacting particle systems}
\label{mappinglin}
The evolution equation, Eq. \ref{lin1}, in addition to being a linear approximation to the KPZ equation
with a drift term, also arises in a number of microscopic 
interacting particle systems. Examples are the Katz-Lebowitz-Spohn (KLS)
model at a specified value of the
temperature, and the Asymmetric Random Average Process (ARAP). These correspondences are
discussed below.
\subsection{Correspondence to the KLS model}
The aim of this subsection is to explain how the variance of the 
displacement, computed in the linearized continuum theory and defined in
terms of the macroscopic parameters ($\Gamma$, $A$, speed $u$),
compares to the same quantity evaluated by numerical simulations in the
(discrete) KLS model, which uses one single microscopic parameter,
namely, the temperature, for a special value of the latter.

\subsubsection{KLS model: a reminder}

The one-dimensional KLS model generalizes the ASEP, in that interactions
between particles are added on top of the exclusion constraint
\cite{KLS}. The ASEP is the infinite temperature limit of the KLS model.

The precise definition of the KLS model is as follows.
Consider a chain of Ising spins $s_{n}$ ($n=1,\ldots,N$), evolving under
the Kawasaki dynamics with the heat-bath rule, and submitted to a drift.
The energy of the chain reads
\begin{equation}
{E}=-\tilde{J}\sum_{n}s_{n}s_{n+1}.
\label{ham}
\end{equation}

In the heat-bath dynamics, for a pair of opposite spins
($s_{n}+s_{n+1}=0$), the move
$(s_{n}\to-s_{n},s_{n+1}\to-s_{n+1})$ is realized
with probability 
\begin{equation} 
W(\Delta E)=\frac{\cal P}{e^{\beta\Delta E}+1}, 
\label{heat} 
\end{equation}
where $\Delta E=2\tilde{J}(s_{n-1}s_{n}+s_{n+1}s_{n+2}) =0,\pm
4\tilde{J}$ is the energy
difference between the configurations after and before the move, and the
numerator $\cal P$ is taken equal to $p$ if the $+$ spin is exchanged to
the right (i.e., $+-\to-+$), and to $q$ if it is exchanged to the left
($-+\to+-$).  Associating a particle to a $+$ spin, and a hole to a $-$
spin, this particle hops to the right with probability $p$, and to the
left with probability $q$.

The moves corresponding to the three possible values
of the difference $\Delta E$ are listed in the Table \ref{movsym},
with the corresponding acceptance probabilities.
Evaporation corresponds to the detachment of a $+$ spin from a positive
domain, or, equivalently, to the detachment of a particle from a cluster of particles.
Condensation, conversely, corresponds to the attachment of an isolated $+$ spin to
a positive domain, or, equivalently, to that of a particle to a cluster of particles, and the two diffusion mechanisms, to the motion of 
an isolated $-$ spin in a positive domain (or hole in a cluster of particles) or to that of an isolated $+$ spin in a negative domain (or particle amongst empty sites).

The heat-bath rule, Eq. \ref{heat}, has the non-trivial property that the steady state is independent
of the asymmetry~\cite{KLS,lg}.
It obeys detailed balance with respect to the energy, Eq. \ref{ham}, at
temperature $T=1/(k_{B}\beta)$ ($k_{B}$: Boltzmann constant),
i.e., $W(\Delta E)=W(-\Delta E)e^{-\beta\Delta E}$.
\begin{table}[ht]
\begin{center}
\begin{tabular}{|l|c|c|c|}
\hline
Type&$\Delta E$&Acceptance~Prob.&Moves\\
\hline
Condensation&$-4\tilde{J}$&${\cal P}\,e^{4\beta \tilde{J}}/(e^{4\beta
\tilde{J}}+1)$
&$\matrix{-+-\,+\to--+\,+\cr+-+\,-\to++-\,-}$\\
\hline
Diffusion&0&${\cal
P}/2$&$\matrix{++-\,+\leftrightarrow+-+\,+\cr--+\,-\leftrightarrow-+-\,-}$\\
\hline
Evaporation&$+4\tilde{J}$&${\cal P}/(e^{4\beta \tilde{J}}+1)$
&$\matrix{++-\,-\to+-+\,-\cr--+\,+\to-+-\,+}$\\
\hline
\end{tabular}
\caption{\small Types of moves in the partially asymmetric Kawasaki dynamics,
and corresponding acceptance probabilities with the heat-bath rule.
The probability ${\cal P}$ is equal to $p$ when the $+$ spin exchanges to the right, to $q$ when it exchanges to the left.}
\label{movsym}
\end{center}
\end{table}

The KLS model can also be viewed as a migration process (or urn model).
Generally speaking, a migration process 
can be seen as a generalization of the Zero Range Process (ZRP)
\cite{zrp,review}, where the particles hop on a lattice,
but where the hopping rate depends both on the departure and arrival
sites [35, 34]. %~\cite{kelly,review}.
Particles (or $+$ spins) are identified with the sites of a lattice, and 
the holes (or $-$ spins) with particles located on these sites.
At infinite temperature, this maps the ASEP onto the usual ZRP.
At finite temperature, the rate at which a particle hops to the neighboring
site is still given by Eq. \ref{heat} with $\Delta E=-4 \delta \tilde{J} $ where $\delta$ is the variation of the number of empty sites before and after moving the particle.
This mapping is effectively used in numerical simulations. 

\subsubsection{The variance of the displacement}

The displacement of a tagged particle in the KLS model
obeys the same laws as for the ASEP, i.e., all phenomena presented in
this paper for
the ASEP exist likewise for the KLS model.
More precisely, the temporal behavior of the variance of the displacement
of the tagged particle is the same as for the ASEP, up to prefactors which depend continuously on the temperature. 
The question is now to make a link between the microscopic model and its
continuum limit description, i.e., to compare the prediction of Section
\ref{exactsol} for the temporal evolution of the variance, in the linear theory, to numerical simulations of the model.

For that purpose, we will (i) express the parameters of the continuum
theory ($\Gamma$, $A$, and $u$) in terms of the temperature, which is the only parameter of the KLS model, (ii) explain the relevance of the linear theory for a special value of the temperature, for which the coefficient of the nonlinear term in the KPZ equation vanishes.
This identification is possible in the present case because the stationary state of the KLS model is known.
The method is given in \cite{gl}. It needs to be slightly generalized
here, as now explained. For the sake of simplicity, we will restrict the
computations to the case of zero magnetization $M$, or, equivalently, of
density $\rho=1/2$ (since $M=2\rho-1$). 

First, the following relationship between the diffusion constant $\Gamma$ and the strength of the noise $A$, is always valid, independently of the
asymmetry \cite{gl},
\begin{equation}
A=\Gamma\,\chi,
\label{Adef}
\end{equation}
where $\chi$ is the susceptibility, explicitly known
for the Ising chain, $\chi=e^{2\beta \tilde{J}}$.
In order to find an expression of $\Gamma$ and of the speed of the tagged particle, we write the equation for the temporal evolution of the magnetization $\langle s_n\rangle$. This reads
\begin{equation}
\label{master}
\frac{d \langle s_{n}\rangle}{d t}= J^{sp}_{n}-J^{sp}_{n+1},
\end{equation}
where the spin current $J^{sp}_{n}$ through the link $(n-1,n)$ has the following
expression.
\begin{widetext}
\begin{eqnarray}
J^{sp}_{n}&=&\frac{p-q}{4}\left (
1-\langle s_{n-1}s_{n}\rangle+
\frac{e^{4\beta \tilde{J}}-1}{2(e^{4\beta \tilde{J}}+1)}\langle
(s_{n}-s_{n-1})(s_{n-2}-s_{n+1})\rangle
\right )\nonumber\\
&+&\frac{1}{4} \Big (
\langle s_{n-1}\rangle-\langle s_{n}\rangle
+\frac{e^{4\beta \tilde{J}}-1}{2(e^{4\beta \tilde{J}}+1)}
(\langle s_{n+1}\rangle-\langle s_{n-2}\rangle +\langle s_{n-2}s_{n-1}s_{n}\rangle
-
\langle s_{n-1}s_{n}s_{n+1}\rangle)\Big ).
\label{current}
\end{eqnarray}
\end{widetext}
The first line in Eq. \ref{current} is the drift term. It is only present when the dynamics is asymmetric. 
It does not contribute to the computation of the diffusion constant
$\Gamma$. Putting aside the drift term, the right-hand side of Eq.
\ref{current} appears as a second order difference, consistently with the diffusive nature of the second term.
(Note also that the first term is even in the spin variable, while the second one is odd.)
In other words, the expression of $\Gamma$ is the same both in the symmetric (EW) or asymmetric (KPZ) cases. It reads (see~\cite{gl} for the computation)
\begin{equation}
\Gamma=\frac{1}{(e^{2\beta \tilde{J}}+1)(e^{4\beta \tilde{J}}+1)}.
\end{equation}
Note that this expression differs by a factor 2 with that given in~\cite{gl}.
This is due to the difference in the definitions of the scale of time chosen in the present work (see Table~\ref{movsym}), and in~\cite{gl}.
Utilizing Eq. \ref{Adef} and the fact that the susceptibility $\chi=e^{2
\beta \tilde{J}}$, we get
\begin{equation}
A=\frac{1}{(e^{-2\beta \tilde{J}}+1)(e^{4\beta \tilde{J}}+1)}.
\end{equation}

It remains to evaluate the speed $u$, as well as the coefficient of
nonlinearity $\lambda$, appearing in the KPZ equation, Eq.
\ref{kpzdrift}. We use Eqns. \ref{u} and \ref{lambda} for this purpose.
Now, without
loss of generality, we consider completely asymmetric particle (i.e.,
$+$ spin) motion with $p=1$ and $q=0$.
For zero magnetization (or, equivalently, at density $\rho=1/2$), since
$u$ is given by $u=\left[v_{P}-\frac{\partial J}{\partial
\rho}\right]\Big |_{\rho=1/2}=\left[v_{P}-2\frac{\partial J}{\partial
M}\right]\Big|_{M=0}$, while $\lambda$ is given by
$\lambda=2\frac{\partial^{2} J}{\partial M^{2}}\Big|_{M=0}$, we need to evaluate the current of particles $J$ as a function of the
magnetization $M$. Note that $J=J^{sp}/2$.
We use Eq. \ref{current}, with $\langle s_{n}\rangle=M$, 
$\langle s_{n} s_{n+1}\rangle=\langle s_{1} s_{2}\rangle$, by translation invariance, etc. Hence the current reads
\begin{equation}
J=\frac{1}{8}\left (1+\frac{e^{4\beta \tilde{J}}-1}{e^{4\beta \tilde{J}}+1}
\langle s_{1}s_{3}\rangle-
\frac{2e^{4\beta \tilde{J}}}{e^{4\beta \tilde{J}}+1}
\langle s_{1}s_{2}\rangle
\right).
\end{equation}
The correlators are known by the transfer matrix method which gives $J$
as \cite{gl}
\begin{widetext}
\begin{equation}
J=\frac{2M^{2}+e^{4\beta \tilde{J}}(1-M^{2})-(1+M^{2})(M^{2}+e^{4\beta
\tilde{J}}(1-M^{2}))^{1/2}}{2(e^{8\beta \tilde{J}}-1)(1-M^{2})}.
\label{currentfin}
\end{equation}
\end{widetext}
Utilizing the last equation, we finally get 
\begin{equation}
u=v_P=2J|_{M=0}=\frac{e^{2\beta \tilde{J}}}{(e^{2\beta \tilde{J}}+1)(e^{4\beta \tilde{J}}+1)}~ ,
\end{equation}
while 
\begin{equation}
\lambda=
\frac{1-3e^{2\beta \tilde{J}}}{e^{2\beta \tilde{J}}(e^{2\beta \tilde{J}}+1)(e^{4\beta \tilde{J}}+1)}.
\end{equation}
Hence for $e^{-2\beta \tilde{J}}=3$ (antiferromagnetic chain), $\lambda$ vanishes.
For such a value of the ``temperature" (actually of $\beta \tilde{J}$),
though the dynamics is asymmetric, the continuum theory is linear.
This yields
\begin{equation}
\Gamma=\frac{27}{40},\quad u=\frac{9}{40},\quad A=\frac{9}{40}.
\label{constkls}
\end{equation}

Note that the unit of time is such that at infinite temperature
\begin{equation}
\Gamma=\frac{1}{4},\quad u=\frac{1}{4},\quad A=\frac{1}{4}.
\end{equation}
These quantities differ by a factor 2 from their expression for the SEP (see Appendix \ref{eqderivation}). As above, the origin of this difference lies in the definitions of the scale of time for these two models (see Table~\ref{movsym}).

\begin{figure}
\begin{center}
\includegraphics[angle=0,scale=0.39]{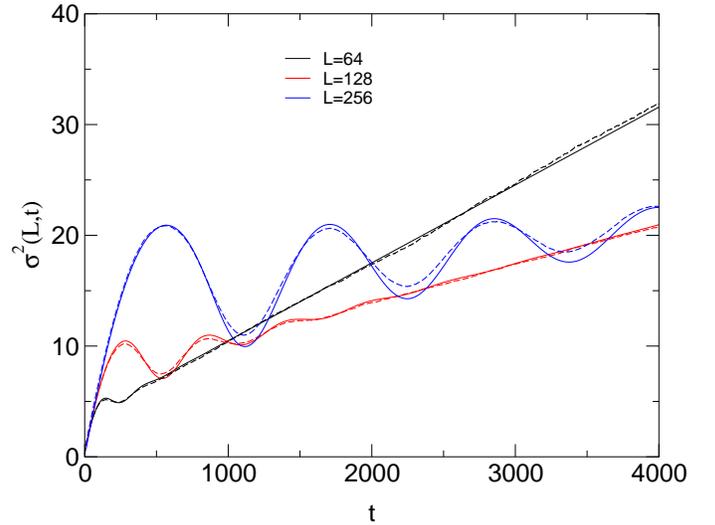}
\caption{(Color online) Comparison of the variance $\sigma^{2}(L,t)$ of
the displacement of the tagged
particle in the KLS model obtained from simulations (dashed
lines) and exact
solution (full lines) of the corresponding continuum linear theory for the
temperature $T$ satisfying $e^{-2\tilde{J}/k_{B}T}=3$. Results are shown for
three different system sizes ($L=64,128,256$). Simulation of the KLS model is done with
the equivalent migration process.}
\label{f1}
\end{center}
\end{figure}

The result of the comparison of the quantity $\sigma^{2}(L,t)$ as
obtained from simulation of the KLS model on a ring of size $L$ for the
temperature $T$ satisfying $e^{-2\beta \tilde{J}}=3$, and that
from the exact solution of the linear interface as given in Eq.
\ref{sigmatlfin} (with the corresponding values of the constants
$\Gamma, A$ and $u$ given in Eq. \ref{constkls}) is shown in Fig.
\ref{f1} for three different system sizes ($L=64, 128, 256$).  
The simulation of the KLS model was done in the equivalent migration
process described above, with equal number of particles and sites, which
corresponds to half filling ($\rho=1/2$) in the original particle system
(i.e. the KLS model). The variance of the displacement of a given tagged
particle in the KLS model translates into the variance of the number of
particles which passed through a given bond in the migration process.

\subsection{Correspondence to the ARAP}
The Asymmetric Random Average Process involves hard core particles
hopping on a real line as opposed to a lattice for the ASEP
\cite{arap}. We
consider a system of particles of average density $\rho$. We denote by
$x_{i}(t)$ the location of the $i$-th particle on the real line with
periodic boundary conditions. The
dynamics is stochastic and involves the following moves during an
infinitesimal time $dt$: a randomly chosen particle jumps with
probability $pdt$ to the right, with probability $qdt$ to the left, and
with probability $1-(p+q)dt$, it continues to occupy its original
location. The amount by which the particle moves either to the right or
to the left is a random fraction of the gap to the next particle to the
right or to the left. Thus, for the $i$-th particle, the jump to the
right is by the amount $r_{i}^{+}(x_{i+1}-x_{i})$, while to the left is
by the amount by $r_{i}^{-}(x_{i}-x_{i-1})$. Here, the random variables
$r_{i}^{\pm}$ are independently drawn from the interval $[0,1]$, each
being distributed according to the same pdf $f(r)$, which is arbitrary.
The time evolution equation of the positions $x_{i}$'s is represented by
the exact Langevin equation
\begin{equation}
x_{i}(t+dt)=x_{i}(t)+\gamma_{i}(t),
\end{equation}
where the random variables $\gamma_{i}(t)$ are given by
\begin{equation}
\gamma_{i}(t)=\left\{\begin{array}{ll}
r_{i}^{+}[x_{i+1}(t)-x_{i}(t)] &\mbox{with prob. $pdt$,} \\
r_{i}^{-}[x_{i-1}(t)-x_{i}(t)] &\mbox{with prob. $qdt$,} \\
0 & \mbox{with prob. $1-(p+q)dt$.}
\end{array}
\right. 
\label{gamma}
\end{equation}
The average velocity of a tagged particle $v_{P}$ is given by
\begin{equation}
v_{P}=\frac{d \langle x_{i}(t) \rangle}{d t}=\langle \gamma_{i}(t) \rangle.
\end{equation}
Using the expression for $\gamma_{i}$ in Eq. \ref{gamma}, we can compute its
average. This gives the average velocity of an ARAP particle as
$ v_{P} =\frac{(p-q)\mu_{1}}{\rho}$, where $\mu_{1}=\int_{0}^{1}dr~r f(r)$.
Thus the average particle current $J=\rho v_{P}$ is independent of
$\rho$. The
derivation of the corresponding interface equation can be done along the
lines outlined in Appendix \ref{eqderivation}. Noting that all
derivatives of $J$ with respect to $\rho$ are zero, Eq. \ref{u}
implies $u=v_{P}$, while $\lambda=0$ from Eq. \ref{lambda}. Thus, we recover the equation
for the linear interface, Eq. \ref{lin1}. Based on our exact solution
for the linear interface, we expect the tagged-particle correlation in
the ARAP in an infinite system as measured by $\sigma^{2}(t)\equiv \lim_{L \rightarrow
\infty}\sigma^{2}(L,t)$ to vary as $t$ (cf. Table
\ref{sigmatable}). This has
been confirmed by an exact solution of the tagged-particle correlation for
the ARAP on an infinite system \cite{arap1}.
\section{Center-of-mass motion}
\label{com}
In this section, we examine the motion of the center-of-mass for
the ASEP, defined as
\begin{equation}
Y_{CM}(t)=\frac{1}{N}\sum\limits_{n=1}^{N}y(n,t).
\end{equation}
Define, for fixed initial configuration, drawn from the stationary
ensemble,
\begin{equation}
d_{CM}(t)=Y_{CM}(t)-Y_{CM}(0)-\langle Y_{CM}(t)-Y_{CM}(0)\rangle,
\end{equation}
where, as usual, angular brackets denote averaging over noise in
the evolution of configurations.
On the basis of the representation in Eq. \ref{post}, we have
\begin{equation}
y(n,t)-y(n,0)-\langle y(n,t)-y(n,0)\rangle=t^{\beta}[\chi_{n}(t)-\langle \chi_{n}(t) \rangle].
\end{equation}
This gives
\begin{equation}
d_{CM}(t)=\frac{t^{\beta}}{N}\sum_{n=1}^{N}[\chi_{n}(t)-\langle \chi_{n}(t) \rangle]. 
\end{equation}
Hence,
\begin{equation}
\langle d_{CM}^{2}(t) \rangle=\frac{t^{2\beta}}{N^{2}}\sum_{n,m}[\langle
\chi_{n}(t)\chi_{m}(t) \rangle-\langle \chi_{n}(t) \rangle \langle
\chi_{m}(t) \rangle].
\label{zetacorr}
\end{equation}
The bracketed quantity on the right hand side, in the stationary state, 
becomes a function, $f(n-m,t)$, of the difference. Putting this in Eq.
\ref{zetacorr}, we get
\begin{equation}
\langle d_{CM}^{2}(t) \rangle=\frac{t^{2\beta}}{N}\sum_{n-m}f(n-m,t).
\label{dcm2t}
\end{equation}
The random variables $\chi_{n}(t)$ are correlated only up to $\xi(t)
\sim t^{1/z}$. We write
\begin{equation}
\sum_{n-m}f(n-m,t)=\rho \int_{0}^{\xi(t)}dx
~f(x,t)\sim \rho  t^{1/z}\int_{0}^{1}dy ~g(y).
\label{fnmt}
\end{equation}
In arriving at the second step in the above equation, we have made a
transformation of the tag variable $(n-m)$ to the spatial variable $x$.
Plugging Eq. \ref{fnmt} in Eq. \ref{dcm2t}, we get  
\begin{equation}
\langle d_{CM}^{2}(t) \rangle \sim \frac{t^{2\beta+1/z}}{L}.
\label{dcm2t2}
\end{equation}
The above result is derived on the basis of the representation for the tagged
particle displacement, Eq. \ref{post}, based on a scaling argument. Using the mapping in Table
\ref{mapping}, Eq. \ref{dcm2t2} is
expected to hold for an interface with nonequilibrium dynamics with
critical exponents $\beta$ and $z$ for times $t \ll L^{z}$. For large times $t\gg L^{z}$, the
fluctuations are expected to grow diffusively, as indicated by the
linear behavior of the variances $\sigma^{2}(L,t)$ and $s^{2}(L,t)$. Thus, for $t \gg
L^{z}$, we expect $\langle d_{CM}^{2}(t) \rangle \sim D(L)t$. Matching the
behavior of $\langle d_{CM}^{2}(t) \rangle$ at the crossover time
$t^{*}\sim L^{z}$, we get $D(L) \sim L^{(2\beta-1)z}$.

With $z=z_{KPZ}=3/2$ and $\beta=\beta_{KPZ}=1/3$ for the ASEP, we have
\begin{equation}
\langle d_{CM}^{2}(t) \rangle \sim \left\{\begin{array}{ll}
            \frac{t^{4/3}}{L}& \mbox{if $t \ll L^{3/2}$} \\
            \frac{t}{\sqrt{L}}& \mbox{if $t \gg L^{3/2}$}.
            \end{array}
            \right.
\end{equation}
The result $\langle d_{CM}^{2}(t) \rangle \sim \frac{t^{4/3}}{L}$ for $t
\ll L^{3/2}$ for the ASEP has already been observed by van Beijeren et al. in \cite{vanb2}.

For the linear interface in the EW class with
$\beta=\beta_{EW}=1/4, z=z_{EW}=2$, we get $D(L) \sim \frac{1}{L}$.
Thus, for the linear interface, we expect, on the basis of our
representation for the tagged particle displacement Eq. \ref{post}, 
\begin{equation}
\langle d_{CM}^{2}(t) \rangle \sim \frac{t}{L}
\label{ewcm}
\end{equation}
for all times. Indeed, the above result, based on scaling arguments, is borne out by an exact
computation of the quantity $\langle d_{CM}^{2}(t) \rangle$ for the linear interface model. Using the mapping of the ASEP to
the interface in Table \ref{mapping}, for an interface of 
length $L$, we get  
\begin{equation}
d_{CM}(t)=\frac{1}{L}\int_{0}^{L}dx~[h(x,t)-h(x,0)-
\langle[h(x,t)-h(x,0)]\rangle].
\label{dcm1}
\end{equation}

We have $\tilde{h}(m,t)=\frac{1}{L}\int_{0}^{L}dx~ h(x,t)e^{-i\frac{2
\pi m}{L}(x+ut)}$.
This gives $\tilde{h}(0,t)=\frac{1}{L}\int_{0}^{L}dx~ h(x,t)$.
Substituting in Eq. \ref{dcm1}, we get
\begin{equation}
d_{CM}(t)=\tilde{h}(0,t)-\tilde{h}(0,0)-\langle [\tilde{h}(0,t)-
\tilde{h}(0,0)] \rangle.
\end{equation}
Using Eq. \ref{htfour}, we finally get
\begin{equation}
d_{CM}(t)=\int_{0}^{t}dt' \tilde{\eta}(0,t')
\end{equation}
so that
\begin{eqnarray}
\langle d_{CM}^{2}(t) \rangle&=&\int_{0}^{t}\int_{0}^{t}dt' dt''
\langle \tilde{\eta}(0,t')\tilde{\eta}(0,t'') \rangle \nonumber \\
&=&\frac{2At}{L},
\label{exdcm2t}
\end{eqnarray}
where we have used Eq. \ref{etacorr} in arriving at the last
equation. This exact result matches with the one obtained on the basis
of scaling arguments in Eq. \ref{ewcm}.

\section{Conclusion}
\label{summary}
In this paper, we have revisited the problem of tagged particle
correlation in the ASEP in one dimension, with particular emphasis on finite size effects. The stationary
state variance of the displacement of a tagged particle involves 
averaging with respect to an initial stationary ensemble and 
stochastic evolution. This quantity shows two distinct size-dependent
time scales, $T_{1} \sim L$, and $T_{2} \sim L^{3/2}$.
The behavior is linear for both the time regimes $t \ll T_{1}$ and 
$t \gg T_{2}$, while for intermediate times $T_{1} \ll t \ll T_{2}$,
the variance $\sigma^{2}(L,t)$ shows pronounced oscillations with a well-defined
size-dependent time period. We understand the oscillations as 
arising from the sliding density fluctuations (SDF) relative to the drift
of the tagged particle in the stationary state, the density
fluctuations themselves being transported through the system by
kinematic waves. 

Following van Beijeren \cite{vanb1}, we also study the variance
$s^{2}(L,t)$ of the displacement of the tagged particle by averaging with
respect to only stochastic evolution of a fixed initial
configuration, drawn from the ensemble of stationary states. We see
that the quantity $s^{2}(L,t)$ has only one time scale $T^{*} \sim
L^{3/2}$. The variance $s^{2}(L,t)$ grows linearly in time for times 
$t \gg T^{*}$, while for times $t \ll T^{*}$, it grows as $t^{2/3}$.

Through a mapping to an interface, the time evolution equation
for the stationary state ASEP density profile maps onto that for a
nonequilibrium interface in the Kardar-Parisi-Zhang (KPZ) universality
class of interface dynamics. This equation is nonlinear and cannot
be solved exactly. By dropping the nonlinear term, we obtain a linear
model in the Edwards-Wilkinson (EW) universality class, which we
solve exactly. This exact solution was helpful in understanding the
occurrence of size-dependent time-scales $T_{1}$, $T_{2}$ and $T^{*}$,
besides explaining the scaling properties of two other interacting
particle systems.

\section{Acknowledgments}
We acknowledge useful discussions with H. van Beijeren, G.
Sch\"{u}tz and J. M. Luck. We are grateful to the 
Isaac Newton Institute, Cambridge, UK, where many of these discussions
took place during the workshop ~`Principles of
the Dynamics of Non-Equilibrium Systems', for its warm hospitality. MB's stay at the Newton 
Institute was supported through EPSRC Grant 531174. MB and SNM
acknowledge the support of the Indo-French Centre for the Promotion of
Advanced Research (IFCPAR) under Project 3404-2. SG is grateful
to the Kanwal Rekhi Career Development Fund for partial financial
support.
\appendix
\section{A primer on kinematic waves}
\label{kinwaves}
The notion of kinematic waves goes back to the work of Lighthill 
and Whitham, who showed how their occurrence follows from the
continuity equation. We briefly recapitulate their argument
\cite{lighthill1,whitham} here.

The steady state coarse-grained density fluctuation, $\delta 
\rho=\rho -\langle \rho \rangle$, satisfies the equation of continuity
\begin{equation}
\frac{\partial (\delta \rho)}{\partial t}=-\frac{\partial j
}{\partial x},
\label{cont1}
\end{equation}
where $j$ is the particle current. Assuming that $j$ depends on
$x$ through its dependence on a space-varying density $\rho$, the
above equation can be recast into
\begin{equation}
\frac{\partial \rho}{\partial t}=-c(\rho)\frac{\partial \rho}
{\partial x},
\label{cont2}
\end{equation}
where $c(\rho)=\frac{\partial j}{\partial \rho}$. To lowest order
in $\delta \rho$, Eq. \ref{cont2} gives
\begin{equation}
\frac{\partial \rho}{\partial t}=-v_{K}\frac{\partial \rho}
{\partial x}
\label{cont3}
\end{equation}
with $v_{K}=c(\langle \rho \rangle)$. To leading order in 
$\delta \rho$, we have $v_{K}=\frac{\partial J}{\partial 
\rho}$, where $J$ is the mean current in the steady state. Equation 
\ref{cont3} is the familiar wave equation in one dimension, and the 
solution reads $\delta \rho=g(x-v_{K}t)$. Here, $g$ is an arbitrary
function of its argument. Thus, to leading order in $\delta \rho$,
the solution to the equation of continuity gives kinematic waves of 
velocity $v_{K}$ that transport density fluctuations through the system
in the stationary state. The attribute `kinematic' emphasizes the
purely kinematic origin of these waves, in contrast to the dynamic 
origin of say, elastic and acoustic waves. $v_{K}$ is identical to 
the collective velocity discussed in \cite{schutz}. Note that this
description neglects higher order terms in $\delta \rho$, which is
tantamount to neglecting dissipation in the density profile while it
is being bodily transported by the kinematic waves.
\section{Derivation of the interface equation}
\label{eqderivation}
Here, we derive the interface equation, Eq. \ref{kpzdrift}, 
following \cite{satya1}.

The local interparticle distance in the ASEP is given by
\begin{equation}
\frac{1}{\rho(x,t)}=\frac{\partial y}{\partial x},
\end{equation}
where $y(x,t)$ is the position of the $x$-th particle at time $t$.
Now,
\begin{equation}
y(x,t)=\frac{x}{\rho}+h(x,t),
\end{equation}
where $\rho$ is the mean density of particles and $h(x,t)$ is the
displacement of the $x$-th particle from the position it would have had, had the particles been uniformly placed.
Then, we have $\rho(x,t)^{-1}=\rho^{-1}+\frac{\partial h}{\partial x}$, implying
\begin{equation}
\rho(x,t)=\frac{\rho}{1+\rho \frac{\partial h}{\partial x}}.        
\end{equation} 

Expanding in a power series in $\frac{\partial h}{\partial x}$,
\begin{equation}
\rho(x,t)=\rho-\rho^{2}\frac{\partial h}{\partial x}+\rho^{3}
\left(\frac{\partial h}{\partial x}\right)^{2}+\ldots
\end{equation}
The above equation can be rewritten as 
\begin{equation}
 \rho(x,t)=\rho+\psi(x,t) ~~\mathrm{where} ~~\psi(x,t)=-\rho^{2}\frac{\partial h}
 {\partial x}+\rho^{3}\left(\frac{\partial h}{\partial x}\right)^{2}
 +\ldots
\end{equation}
In the absence of any drift velocity, the equation of motion of
$h(x,t)$ is diffusive.
\begin{equation}
\frac{\partial h}{\partial t}=\Gamma'\frac {\partial^{2}h}{\partial x^{2}}+\eta(x,t).
\end{equation}
The noise term $\eta(x,t)$ is Gaussian: $\langle \eta(x,t) \rangle=0, \langle \eta(x,t) \eta(x',t') 
\rangle = 2A \delta(x-x') \delta(t-t')$. 

In the presence of a drift velocity $v(x,t)$, an additional
term $v(x,t)$ appears on the rhs of the above equation. The drift
velocity depends on $x$ and $t$ only through the local density
$\rho(x,t)$ which is the only possibility in the coarse-grained 
lattice gas. Thus $v(x,t) = v(\rho(x,t))$. We now substitute 
$\rho(x,t)=\rho+\psi(x,t)$ and expand in a power series in
$\psi(x,t)$.  
\begin{equation}
v(x,t)=v(\rho)+\left[\frac{\partial v}{\partial \rho}\right]_
{\rho}\psi(x,t)+\frac{1}{2}\left[\frac{\partial^{2}v}{\partial
\rho^{2}}\right]_{\rho}\psi^{2}(x,t)+\ldots
\end{equation}
Thus, we get
\begin{eqnarray}
\frac{\partial h}{\partial t}&=&v(\rho)+\Gamma'\frac{\partial ^{2}h}
{\partial x^{2}}-\rho^{2}\left[\frac{\partial v}{\partial \rho}\right]_
{\rho}\frac{\partial h}{\partial x}+\frac{1}{2}
\left[\frac{\partial ^{2}v}{\partial \rho^{2}}\right]_
{\rho}\rho^{4}\left(\frac{\partial h}{\partial x}\right)^{2}
\nonumber \\
&+&\left[\frac{\partial v}{\partial \rho}\right]_{\rho}\rho^{3}
\left(\frac{\partial h}{\partial x}\right)^{2}+\ldots+\eta(x,t) \\
          &=&v(\rho)+\Gamma'\frac{\partial ^{2}h}{\partial x^{2}}+u'
		\frac{\partial h}{\partial x}+\frac{\lambda'}{2}\left(\frac{\partial h}
		{\partial x}\right)^{2}+\ldots+\eta(x,t). \nonumber \\
\label{intereqn}
\end{eqnarray}
Here, $v(\rho)$ is the mean drift velocity of the particle and hence,
equals $v_{P}$ in our notation. Also, $u'=\rho\left[v(\rho)-\frac{\partial J}{\partial
\rho}\right]_{\rho}$ where $J(\rho)=\rho v(\rho)$ is the mean current.
In terms of the kinematic wave speed $v_{K}$, we have
$u'=\rho(v_{P}-v_{K})$. The nonlinearity coefficient
$\lambda'=\rho^{3}\left[\frac{\partial^{2}J}{\partial
\rho^{2}}\right]_{\rho}$. Note that $x$ in the above equation stands for
the tag variable in the continuum. 
Now, dividing $x$ by the particle density $\rho$ to make it into a
spatial variable, we finally get from Eq. \ref{intereqn}, to lowest order of
nonlinearity,
\begin{equation}
\frac{\partial h}{\partial t}=v_{P}+\Gamma\frac{\partial^{2}h}{\partial x^{2}}+u
		\frac{\partial h}{\partial x}+\frac{\lambda}{2}\left(\frac{\partial h}
		{\partial x}\right)^{2}+\eta(x,t),
\label{intereqn1}                
\end{equation}
where 
\begin{equation}
u=v_{P}-v_{K},
\label{u}
\end{equation}
while
\begin{equation}
\lambda=\rho\left[\frac{\partial^{2}J}{\partial \rho^{2}}\right]_{\rho}.
\label{lambda}
\end{equation}
Using $J(\rho)=(p-q)\rho(1-\rho)$ and $v_{P}=(p-q)(1-\rho)$ for the ASEP, we get
\begin{eqnarray}
u&=&\rho(p-q), \\
\lambda&=&-2\rho(p-q).
\end{eqnarray}
Utilizing the above expressions for $u$ and $\lambda$ in Eq.
\ref{intereqn1}, we get the time evolution
equation for the interface equivalent to the ASEP, Eq. \ref{kpzdrift}.

Expressions for the coefficients $\Gamma$ and $A$ in terms of
microscopic parameters can be found by setting $p=q=1/2$, in which case
the ASEP reduces to the SEP. The coefficient $\Gamma$ can then be
calculated explicitly \cite{kutner, stinchcombe}, with the result
\begin{equation}
\Gamma=\frac{1}{2}.
\label{gammaderv}
\end{equation}
Further, Eq. \ref{intereqn1} reduces to the EW equation, Eq. \ref{ew},
and $\sigma^{2}(t)$ can be found from Section \ref{exactsol} by
considering the limit $u \rightarrow 0$. The result is $\sigma^{2}(t)
\approx \frac{2A}{\sqrt{\pi \Gamma}} \sqrt{t}$. Comparing with the exact
result $\sigma^{2}(t) \approx \sqrt{\frac{2}{\pi}}
\left(\frac{1-\rho}{\rho}\right) \sqrt{t}$ for the SEP \cite{ligget}, one gets
$\frac{A}{\sqrt{\Gamma}}=\frac{1}{\sqrt{2}}\left(\frac{1-\rho}{\rho}\right)$.
Using Eq. \ref{gammaderv}, we finally get
\begin{equation}
A=\frac{1}{2}\left(\frac{1-\rho}{\rho}\right).
\end{equation}

\section{Evaluation of the integral $\int_{k=0}^{\infty}
\frac{dk}{k^{2}}[1-e^{-ck^{2}}\cos(k)]$}
\label{integralev}
Let 
\begin{equation}
I(c)=\int_{k=0}^{\infty}\frac{dk}{k^{2}}[1-e^{-ck^{2}}\cos(k)].
\end{equation}
Thus,
\begin{equation}
\frac{dI}{dc}=\int_{k=0}^{\infty}dk~e^{-ck^{2}}\cos(k)=\frac{1}{2}\sqrt
{\frac{\pi}{c}}e^{-1/4c}.
\end{equation}
Also, $I(0)=\frac{\pi}{2}$. Hence,
\begin{equation}
I(c)=\frac{\pi}{2}+\frac{1}{2}\int_{0}^{c} dx \sqrt{\frac{\pi}{x}}e^{-1/4x}.
\end{equation}
Doing the integral on the rhs by parts, we finally get
\begin{equation}
I(c)=\frac{\pi}{2}+\sqrt{\pi c}~e^{-1/4c}-\frac{\sqrt{\pi}}{2}
\int_{1/4c}^{\infty}dy ~e^{-y}y^{-1/2}.
\end{equation}
Using the usual definition of the complementary error function,
$\mathrm{erfc}(z)=\frac{2}{\sqrt{\pi}}\int_{z}^{\infty} dt
~e^{-t^{2}}=1-\mathrm{erf}(z)$, (where $\mathrm{erf}(z)$ is the usual error function), we can rewrite the above expression as
\begin{eqnarray}
I(c)&=&\frac{\pi}{2}+\sqrt{\pi c}~e^{-1/4c}-\frac{\pi}{2} ~\mathrm{erfc}
\left(\frac{1}{2\sqrt{c}}\right) \\ \nonumber
&=&\sqrt{\pi c}~e^{-1/4c}+\frac{\pi}{2}~\mathrm{erf}\left(\frac{1}{2\sqrt{c}}\right).
\end{eqnarray}

\end{document}